\documentclass[article]{aa}
\usepackage{graphicx}
\usepackage[varg]{txfonts}
\usepackage[dvipsnames]{xcolor}
\usepackage{hyperref}
\hypersetup{pdfborder = {0 0 0},
    colorlinks = true,
    linkbordercolor = {white},
    citecolor=NavyBlue,
    urlcolor=Green
}

\newcommand{\orcidlink}[1]{\protect\href{https://orcid.org/#1}{\protect\includegraphics[width=8pt]{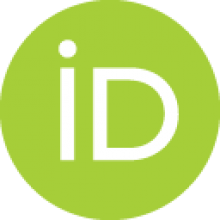}}}
\usepackage{makecell}
\usepackage{amssymb}
\usepackage{amsmath}
\usepackage{xspace}
\usepackage{epstopdf}
\usepackage{rotating}
\usepackage{array,multirow}
\usepackage{enumitem}
\usepackage[english]{babel}
\usepackage{mathrsfs}
\usepackage{listings}
\usepackage{blindtext}
\usepackage{lipsum} 
\usepackage{subfigure}
\usepackage[T1]{fontenc}
\usepackage[utf8]{inputenc}
\usepackage{ae,aecompl}
\usepackage{caption}
\usepackage{threeparttable}
\usepackage{booktabs}
\usepackage{longtable}
\usepackage{natbib,twoopt}
\usepackage{siunitx}

\renewcommand{\eqref}[1]{Eq.~\ref{#1}}

\makeatletter
\def\maketag@@@#1{\hbox{\m@th\normalfont\normalsize#1}}
\makeatother

\bibpunct{(}{)}{;}{a}{}{,} 

\makeatletter
\newcommand*\mysize{%
  \@setfontsize\mysize{5.7}{8.0}%
}
\makeatother

\makeatletter
\newcommand*\tabsize{%
  \@setfontsize\tabsize{7.}{8.0}%
}
\makeatother

\makeatletter
\newcommand\footnoteref[1]{\protected@xdef\@thefnmark{\ref{#1}}\@footnotemark}
\makeatother

\begin{document}
\title{Radial velocity follow-up of \textit{Gaia} astrometric substellar companions: \\ six confirmed brown dwarfs and 13 impostor binaries}

   \titlerunning{RV follow-up of \textit{Gaia} astrometric substellar candidates}

   \author{
    Marcus L. Marcussen\inst{\ref{I1}}\orcidlink{0000-0003-2173-0689}\corrauth{marcus@phys.au.dk} \and
    Simon H.\ Albrecht \inst{\ref{I1}}\orcidlink{0000-0003-1762-8235}   \and
    Kamil K. Kalinowski \inst{\ref{I1}}\orcidlink{0000-0001-9825-7418}\and
    Joshua N. Winn\inst{\ref{I5}}\orcidlink{0000-0002-4265-047X} \and
    Guðmundur~Stefánsson \inst{\ref{I2},\ref{I3}}\orcidlink{0000-0001-7409-5688} \and
    Jens Reersted Larsen \inst{\ref{I1}}\orcidlink{0009-0006-0423-2353} \and
    Julie Gadeberg\inst{\ref{I7}} \and
    Evan Fitzmaurice\inst{\ref{I8},\ref{I4}}\orcidlink{0000-0003-0199-9699} \and
    Kevin~Schlaufman\inst{\ref{I6}}\orcidlink{0000-0001-5761-6779} \and 
    Suvrath Mahadevan\inst{\ref{I2},\ref{I4}}\orcidlink{0000-0001-9596-7983}
    }

   \institute{Stellar Astrophysics Centre, Institut for Fysik og Astronomi, Aarhus Universitet, Ny Munkegade 120, 8000 Aarhus C, Denmark\label{I1} \and
   Princeton University, 4 Ivy Lane, Princeton, NJ 08540, USA\label{I5} \and
   Astrophysics \& Space Center, Schmidt Sciences, New York, NY 10011, USA \label{I2} \and
   Anton Pannekoek Instituut voor Sterrenkunde, Universiteit van Amsterdam, Science Park 904, 1098 XH Amsterdam, The Netherlands \label{I3} \and
   Center for Exoplanets and Habitable Worlds, 525 Davey Laboratory, The Pennsylvania State University, University Park, PA 16802, USA\label{I4}  \and
   William H.\ Miller III Department of Physics \& Astronomy, Johns Hopkins University, 3400 N Charles St, Baltimore, MD 21218, USA\label{I6} \and
   Nordic Optical Telescope, Rambla José Ana Fernández Pérez 7, Breña Baja, La Palma 38711, Spain\label{I7} \and
   Department of Astronomy \& Astrophysics, The Pennsylvania State University, 525 Davey Laboratory, University Park, PA 16802, USA\label{I8}}
   \date{Received 1 May 2026 / Accepted dd September 2026}
   \authorrunning{Marcussen et al.}

\abstract
{With microarcsecond precision, \textit{Gaia} has made the astrometric discovery of substellar companions feasible. However, follow-up observations
are needed to check on orbital solutions and rule out astrophysical false positives. We validate and characterise a sample of 20 \textit{Gaia} Data Release 3 (DR3) astrometric candidates for substellar companions, with the dual goal of identifying false positives and deriving robust physical parameters for genuine companions. We performed high-resolution spectroscopy using FIES, NIRPS, and NEID at the Nordic Optical Telescope, the ESO 3.6m telescope, and the WIYN Telescope, respectively. Double-lined spectroscopic binaries were identified as false positives, and surviving systems were characterised through joint \textit{Gaia} astrometry and radial velocity modelling. Thirteen of the candidates proved to be near-twin binary stars for which the astrometric
motion of the center of light is small
enough to mimic that of a single star
in response to a substellar companion.
We confirm that six companions are brown dwarfs with masses in the range ${\sim}26$--$68\,M_\mathrm{Jup}$, two of them for the first time. In contrast to earlier studies of lower-mass \textit{Gaia} candidates, we find that the
astrometric orbital solutions are generally
reliable to within the stated uncertainties.}

   \maketitle
   \nolinenumbers

\section{Introduction}
\label{sec:introduction}
The detection of unseen companions through their gravitational influence on visible stars is one of the oldest techniques in astrophysics. Long before the first exoplanet was discovered, Bessel famously inferred the existence of Sirius B and Procyon B solely from the anomalous proper motions of their primaries \citep{Bessel1844}. The best positional accuracies of Bessel’s era were around 1\,arcsec, whereas a typical astrometric epoch measurement from the \textit{Gaia} satellite \citep{Prusti+2016} reaches about 50\,$\mu$as \citep{Lindegren2021}, bringing the astrometric discovery of substellar companions within observational reach. In this work, we use “substellar” to refer to companions below the hydrogen-burning minimum mass, which is close to, but below, $80\,M_{\rm Jup}$ for solar composition \citep[e.g.][]{Chabrier2023}. 

With its microarcsecond precision, long time baseline, and all-sky coverage, \textit{Gaia} is expected to transform the census of wide-orbiting giant planets and brown dwarfs. Predicted yields for the end-of-mission data release extend up to tens of thousands \citep{Perryman2014,El-Badry2024,Lammers2026}.

The substellar companion candidates emerging from \textit{Gaia} Data Release 3 (DR3), based on ${\sim}33$ months of data, have yielded both notable successes and challenging cases. Comparing DR3 orbital solutions to ground-based radial velocity (RV) constraints, \citet{Winn2022} found substantial discrepancies for four out of seven planetary candidates. One of the candidates, HIP\,66074\,b, was later confirmed by \citet{Sozzetti2023}
and designated Gaia-3\,b, after an underestimation of the eccentricity was identified in the originally published astrometric solution.
A similar case was presented by \citet{Pinamonti2026}: the DR3 astrometric
solution for Gaia-6\,b featured a short-period, low-eccentricity companion, but the RV
data revealed a longer period and a higher eccentricity. \citet{Fitzmaurice2024} found similar tensions in orbital solutions extending to more massive substellar candidates. \citet{Unger2023} performed joint fits of archival RVs and the DR3 orbital solutions for 31 systems hosting validated substellar candidates, reclassifying 13 companions as low-mass stars. They found good agreement between the \textit{Gaia} solutions and the RVs in most cases, but identified significant discrepancies for individual systems.

Recognizing errors in orbital solutions is not the only challenge. Many \textit{Gaia} substellar candidates have proven to be astrophysical false positives, most frequently
binary stars in which the components
have nearly equal masses and luminosities.
These near-twin binaries
can masquerade as stars with substellar companions
when they are too distant to resolve
both components \citep{Vandekamp1975,Holl2023}. In that case, \textit{Gaia} tracks the photocentre of the unresolved binary, whose motion is quite small when the two stars
bear similar fluxes and move in opposite directions.
Consistent with this picture, \citet{MarcussenAlbrecht2023} found that the three \textit{Gaia} DR3 substellar candidates that they observed are near-twin binaries, and \citet{Stefansson2025} found the same for 21 of their 28 targets. However, \citet{Stefansson2025} 
also confirmed Gaia-4b and Gaia-5b to be genuine substellar companions. \citet{Lammers2026} simulated the population of astrometric
planets and false positives and predicted that the false-positive rate will decline as it becomes possible to detect companions in wider orbits. This photocentre degeneracy can also be addressed using the system photometry: because stellar luminosity depends on mass, the observed \textit{Gaia} photometry can constrain the component flux ratio and thereby help distinguish near-equal-mass stellar binaries from star--substellar systems. \citet{Bailer2026} demonstrated such an approach using \textit{Gaia} astrometry and three-band photometry to infer component masses for unresolved binaries.

Beyond individual detections, \textit{Gaia} will provide a much clearer view of the architecture and demographics of long-period systems. A large sample of wide-orbit companions, together with the geometric information encoded in their astrometric orbits, will provide a view of long-period system architecture that is rarely accessible by other methods. In multiplanet systems, and in stellar binaries where one component hosts a planet, \textit{Gaia} astrometry can constrain mutual inclinations and thus illuminate the dynamical history of these systems. Initial studies along these lines have already begun with \textit{Gaia} DR3 \citep{Behmard2022,Christian2022}. The scientific reach will expand substantially with \textit{Gaia} Data Releases 4 and 5, which
will provide epoch astrometry rather than only catalogue-level orbital solutions, facilitating joint modelling with external data such as RVs.

In this paper, we report spectroscopic follow-up observations
of 20 \textit{Gaia} DR3 candidates with substellar masses and northern declinations. The spectroscopy served three purposes: to reject binary false positives, to determine host star stellar parameters, and to characterise their substellar companions through a joint analysis of Doppler time series and \textit{Gaia} astrometry.

The paper is organised as follows. Section~\ref{sec:observations} describes the target selection and observations. Section~\ref{sec:methods} presents our derivation of spectroscopic and fundamental stellar parameters (Sect.~\ref{sec:stellar_parameters}), the line-profile diagnostics we used to identify binaries (Sect.~\ref{sec:vetting}), and the joint RV + astrometry orbital modelling (Sect.~\ref{sec:forward}). Section~\ref{sec:results} presents the spectroscopic classification results for the full sample and the orbital solutions for the six confirmed brown dwarfs. Section~\ref{sec:discussion} discusses the parameters of the confirmed brown dwarfs and the implications for \textit{Gaia} astrometric candidate validation.

\section{Data and observations}
\label{sec:observations}

\begin{table*}
\caption{Overview of the observed systems.}
\label{tab:parent_sample_overview}
\centering
\begin{tabular}{l c c c c c c l l}
\hline
\hline
Star & Gaia DR3 & $N_{\mathrm{obs}}$ & $G\,[\mathrm{mag}] $ & $m_2$ [$M_\mathrm{Jup}$] & Exp. time $[\mathrm{s}]$ & $\mathrm{S}/\mathrm{N}$ & Classification & Instrument\\
\hline
HD 5433 & 2778298280881817984 & 10 & 8.53 & 51.9 & 300 & 36.8 & Single-lined & FIES \\
BD+35 228 & 321123400368013696 & 1 & 8.95 & 47.3 & 350 & 28.9 & Binary & FIES \\
LP 769-9 & 5148853253106611200 & 10 & 11.38 & 43.4 & 1600 & 13.9 & Single-lined & FIES \\
& & 15 & & & 652 & 32.4 &  & NIRPS \\
TYC 4338-562-1 & 547770821740800384 & 1 & 9.85 & 52.6 & 650 & 32.7 & Binary & FIES \\
HD 275921 & 230138989266819456 & 3 & 9.96 & 26.3 & 1000 & 42.0 & Binary & FIES \\
PM J04205+8131 & 557717892980808960 & 2 & 12.05 & 8.4 & 1600 & 7.3 & Binary & FIES \\
& & 5 & & & 900 & 3.9 &  & NEID \\
TYC 4730-512-1 & 3205578257602278784 & 1 & 10.60 & 47.2 & 1400 & 31.0 & Binary & FIES \\
HD 30246 & 3309006602007842048 & 21 & 8.14 & 40.0 & 700 & 41.7 & Single-lined & FIES \\
2M0809+07 & 3097799420565217408 & 7 & 11.86 & 39.1 & 1600 & 17.3 & Single-lined & FIES \\
HD 91669 & 3750881083756656128 & 15 & 9.49 & 41.3 & 800 & 26.9 & Single-lined & FIES \\
BPM 89283 & 3665298981300771200 & 1 & 10.68 & 19.7 & 800 & 16.3 & Binary & FIES \\
& & 1 & & & 900 & 30 &  & NEID \\
PM J13580+3141 & 1457486023639239296 & 5 & 11.91 & 13.4 & 1400 & 10.8 & Single-lined & FIES \\
2M1515-08 & 6319620720489927680 & 2 & 12.63 & 23.0 & 1800 & 7.6 & Binary & FIES \\
2M1820+68 & 2259699048817216256 & 5 & 13.14 & 15.6 & 1480 & 5.9 & Binary & FIES \\
L 1143-66 & 1750816651978245248 & 1 & 11.38 & 28.8 & 1500 & 9.2 & Binary & FIES \\
G 261-42 & 2277249663873880576 & 3 & 12.91 & 11.1 & 1480 & 5.9 & Binary & FIES \\
& & 3 & & & 900 & 2.5 &  & NEID \\
LP 341-28 & 1853704144046204032 & 35 & 10.57 & 24.9 & 1100 & 17.2 & Single-lined & FIES \\
Ross 778 & 1739379433242806144 & 1 & 10.74 & 27.8 & 720 & 13.5 & Binary & FIES \\
HD 207740 & 1897143408911208832 & 2 & 7.80 & 32.1 & 379 & 65.0 & Binary & FIES \\
BD+35 5080 & 2879477265016061056 & 2 & 8.96 & 27.1 & 450 & 30.6 & Binary & FIES \\
\hline
\end{tabular}
\tablefoot{$N_{\mathrm{obs}}$ is the number of observations. The classification comes from broadening-function vetting. Exp.\ time is the median exposure time. $\mathrm{S}/\mathrm{N}$ is the median signal-to-noise ratio at 550\,nm (FIES and NEID) and at 1250\,nm (NIRPS). Companion masses $m_2$ assume that the companion is dark.}
\end{table*}

\subsection{Sample}
\label{sec:sample}

Our initial sample of 20 systems is listed in Table~\ref{tab:parent_sample_overview}. The sample was obtained as follows. First, we selected the solutions of type \texttt{astrometric} from the \textit{Gaia} DR3 \texttt{nss\_two\_body\_orbits} table\footnote{\url{https://gea.esac.esa.int/archive/}} \citep{DR3Teaser,DR3Binarystar}. The fitted model of \texttt{astrometric} solutions comprises position, parallax, linear motion, and a single superimposed Keplerian orbit, and is constrained by astrometric data only. From these systems, we selected the brightest targets ($G < 13\,\mathrm{mag}$) with predicted companion masses below $80\,M_{\mathrm{Jup}}$.

Of these 20 initial systems, seven survived our spectroscopic vetting of which six will be described in detail (Gaia-4 was already covered by \citet{Stefansson2025} ). Table~\ref{tab:stellar_params} lists the relevant parameters for these six systems. The derivation of the reported spectroscopic parameters is described in Sect.~\ref{sec:stellar_parameters}, and the photometric and astrometric quantities are taken from \textit{Gaia}'s \texttt{gaia\_source} and \texttt{nss\_two\_body\_orbits} tables. Additionally, we report the quantities renormalised unit weight error (RUWE), significance of the orbital solution, and the number of visibility periods in Table~\ref{tab:stellar_params}, which characterise the robustness of the astrometric detection. The RUWE measures the goodness of fit of the standard five-parameter single-star model and thus quantifies the degree of unmodelled effects, such as Keplerian motion \citep{Lindegren2021}. The number of visibility periods provides an approximate count of the independent astrometric epochs constraining the orbit: \textit{Gaia}'s scanning law often produces clusters of observations within a few days, and because neither the orbital reflex motion nor the scan direction changes much over such short intervals, observations separated by less than four days are combined into a single visibility period. Finally, the `significance' metric is the signal-to-noise ratio of the orbital solution.

\subsection{Literature information on the observed systems}
\label{sec:literature}

Before interpreting the \textit{Gaia} orbital solutions, we searched the literature for previous evidence of multiplicity and companion detections, queried the Washington Double Star catalogue \citep[WDS;][]{Mason2001}, and searched \textit{Gaia} DR3 for resolved companions with consistent parallax and proper motion \citep[see also][]{ElBadry2021,GonzalezPayo2024}.
Previous brown dwarf companion detections have been reported for several systems in our sample. The companion to HD\,91669 was discovered by \citet{Wittenmyer2009} with the Tull spectrograph at McDonald Observatory, while the companion to HD\,5433 was detected with SOPHIE by \citet{Dalal2021}. The companion to HD\,30246 was likewise identified with SOPHIE by \citet{Diaz2012}. All three systems were subsequently revisited by \citet{Unger2023}, who combined the archival RV measurements with the corresponding \textit{Gaia} DR3 orbital solutions.

2MASS J08092840+0731101 (2M0809+07, Gaia DR3 3097799420565217408) has a resolved neighbour at a separation of ${\sim}3.3\,$arcsec. The two stars have similar $G$ magnitudes and consistent \textit{Gaia} DR3 parallaxes and proper motions, and their separation corresponds to a projected separation of ${\sim}600$\,au. Over the 33-month DR3 baseline, the motion of the outer companion is absorbed into the linear proper-motion terms of the astrometric solution. The associated change in velocity, ${\sim}2\,\mathrm{m\,s^{-1}}$, is also far too small to affect the 434-day Keplerian signal. Given the $1.3\,$arcsec diameter of the FIES fibre and typical NOT seeing, some contamination from the neighbour at $3.3\,$arcsec is expected. However, any such contamination is not significant for our analysis because the RV semi-amplitudes considered here are two to three orders of magnitude larger than the contamination-induced signals relevant to high-precision RV work.

Recently, \citet{Barbato2026} reported HARPS-N follow-up observations of 14 \textit{Gaia} DR3 astrometric substellar candidates, five of which overlap with our sample. They confirmed the brown dwarf companion to LP\,341-28 (HIP\,105707) and obtained results for Gaia-4 consistent with the previous confirmation by \citet{Stefansson2025}. In addition, they classified TYC\,4338-562-1, HD\,275921, and TYC\,4730-512-1 as double-lined spectroscopic binaries, in agreement with our classifications.

\subsection{FIES spectroscopy}
\label{sec:fies}

We obtained high-resolution spectra with FIES at the 2.56\,m Nordic Optical Telescope (NOT) \citep{NOT2010}, using the high-resolution mode (resolving power $R\approx 67{,}000$) covering approximately 3700--7300\,\AA. The FIES spectra were automatically reduced and extracted using FIEStool. Because FIES is not temperature stabilised, we tracked instrumental drift by bracketing each science exposure with Th-Ar calibration frames and correcting the drift by linear interpolation.

\subsection{NIRPS spectroscopy}
\label{sec:NIRPS}

For LP 769-9, we also obtained near-infrared spectra with the NIRPS spectrograph \citep{2017Msngr.169...21B, 2017SPIE10400E..18W, 2025A&A...700A..10B} installed at the European Southern Observatory (ESO) 3.6$\,$m telescope at La Silla, under ESO programmes 114.2758, 115.27ZE, 116.296E, and 117.2AJC. The reduced one-dimensional spectra were produced with the ESO NIRPS pipeline (\texttt{nirps} 3.3.0) and cover approximately 9660--19{,}230$\,$\AA\ at a resolving power $R\approx 70{,}000$.

For both FIES and NIRPS, relative RVs were extracted from the reduced spectra using a template-matching technique \texttt{serval} adapted from \citet{Zechmeister2018}. A summary of our FIES and NIRPS observations is presented in Table~\ref{tab:parent_sample_overview}. The complete RV time series are available online.\footnote{ \url{https://www.erda.au.dk/archives/bae3ed22c3a51e23f1a0aaabbddf3066/published-archive.html}.}

\subsection{NEID spectroscopy} \label{sec:neid}
For PM J04205$+$8131, BPM 89283, and G 261-42, we obtained five, one, and three high resolution spectra with the NEID spectrograph. NEID is a temperature-stabilized \citep{stefansson2016,robertson2019} high resolution Doppler spectrograph \citep{schwab2016} on the Wisconsin-Indiana-Yale-NOIRLab (WIYN) 3.5\,m Telescope at Kitt Peak National Observatory in Arizona. The reduced one-dimensional spectra were produced with the NEID Data Reduction Pipeline ({\tt DRP})\footnote{\url{https://neid.ipac.caltech.edu/docs/NEID-DRP/}}, and cover a wavelength range of approximately 3800--9300$\,$\AA\ at a resolving power $R\approx 110{,}000$. We studied both the pipeline radial velocities and the pipeline derived CCFs.

\begin{table*}
\centering
\caption{Summary of stellar and orbital parameters adopted in this work.}
\label{tab:stellar_params}
\small
\centering
\begin{tabular}{l c c c c c c}
\hline\hline
Name & LP 341-28 & HD 91669 & HD 5433 & LP 769-9 & HD 30246 & 2M0809+07 \\
\hline
\multicolumn{7}{l}{\textit{Astrometry}}\\
$\alpha$ (J2016) [deg] & $321.15169006$ & $158.72730825$ & $14.05638798$ & $29.55314164$ & $71.62700948$ & $122.36825883$ \\
$\delta$ (J2016) [deg] & $32.12246953$ & $-13.78873400$ & $15.65600439$ & $-14.88076019$ & $15.47193179$ & $7.51947371$ \\
$\mu_\alpha \cos{\delta}$ [mas\,yr$^{-1}$] & $-156.561 \pm 0.017$ & $116.753 \pm 0.053$ & $48.206 \pm 0.114$ & $181.505 \pm 0.058$ & $87.326 \pm 0.091$ & $-23.928 \pm 0.026$ \\
$\mu_{\delta}$ [mas\,yr$^{-1}$] & $-116.725 \pm 0.027$ & $-176.116 \pm 0.051$ & $-36.319 \pm 0.091$ & $38.046 \pm 0.040$ & $-25.353 \pm 0.067$ & $4.491 \pm 0.017$ \\
$\varpi$ [mas] & $19.755  \pm 0.030$ & $13.925 \pm 0.051$ & $15.764 \pm 0.107$ & $13.801 \pm 0.053$ & $20.501 \pm 0.082$ & $5.432 \pm 0.025$ \\
\hline
\multicolumn{7}{l}{\textit{Orbital solution}}\\
Visibility periods & $26$ & $13$ & $14$ & $18$ & $19$ & $18$ \\
Significance & $36.6$ & $19.34$ & $33.4$ & $16.0$ & $5.7$ & $7.9$ \\
RUWE & $2.819$ & $2.765$ & $4.060$ & $2.747$ & $3.300$ & $1.542$ \\
\hline
\multicolumn{7}{l}{\textit{Photometry}}\\
$G$ [mag] & $10.568$ & $9.491$ & $8.525$ & $11.379$ & $8.139$ & $11.858$ \\
\hline
\multicolumn{7}{l}{\textit{Spectroscopic parameters}}\\
$T_{\rm eff}$ [K] & $4254\pm 70$ & $5396\pm110$ & $5847 \pm 70$ & $4244 \pm 70$ & $5665 \pm 110$ & $5295 \pm 110$ \\
$\log \mathrm{g}$ [cgs] & $4.65 \pm 0.12$ & $4.44 \pm {0.12}$ & $4.40 \pm 0.12$ & $4.66 \pm 0.12$ & $4.38 \pm 0.12$ & $4.47 \pm 0.12$ \\
$\mathrm{[Fe/H]}$ [dex] & $0.10 \pm 0.09$ & $0.27 \pm 0.04$ & $0.02 \pm 0.12$ & $0.05 \pm 0.09$ & $0.16 \pm 0.09$ & $-0.32 \pm 0.09$ \\
\hline
\multicolumn{7}{l}{\textit{Stellar properties}}\\
$M_\star$ [$M_\odot$] & $0.700^{+0.016}_{-0.019}$ & $0.984^{+0.024}_{-0.039}$ & $1.036^{+0.034}_{-0.037}$ & $0.695^{+0.017}_{-0.016}$ & $1.005^{+0.042}_{-0.049}$ & $0.782^{+0.036}_{-0.033}$ \\
$R_\star$ [$R_\odot$] & $0.679^{+0.017}_{-0.014}$ & $0.908^{+0.024}_{-0.017}$ & $1.043^{+0.033}_{-0.032}$ & $0.678^{+0.015}_{-0.013}$ & $1.022^{+0.042}_{-0.033}$ & $0.800^{+0.028}_{-0.027}$ \\
Age [Gyr] & $12.2^{+4.5}_{-6.3}$ & $3.0^{+3.9}_{-2.2}$ & $4.0^{+2.4}_{-1.9}$ & $12.6^{+4.5}_{-6.1}$ & $5.8^{+4.0}_{-3.0}$ & $10.6 \pm 5.6$ \\
\hline
\end{tabular}
\tablefoot{A visibility period is a group of observations separated from other groups by at least four days; the reported value is the number of visibility periods used in the astrometric solution. Significance is the S/N of the astrometric orbital solution. Medians and 16th and 84th percentiles of the posterior probability distributions are reported.}
\end{table*}

\section{Methods}
\label{sec:methods}
The analysis had three stages. We first inferred stellar masses, radii, and ages from the spectroscopic parameters and stellar isochrones. We then used the spectra to seek evidence that each target is a stellar binary rather than a single luminous star with a dark companion, based on the morphology of its broadening function. We then modelled the systems that survived this vetting step with RV-only, astrometry-only, and joint RV + astrometry fits. Below, we describe each step in turn.

\subsection{Stellar parameters}
\label{sec:stellar_parameters}

We obtained the host stars' spectroscopic parameters $T_\mathrm{eff}$, $\log{\mathrm{g}}$, and $[\mathrm{Fe}/\mathrm{H}]$ by matching our observed spectra to a library of well-characterised stars. For this we used \texttt{SpecMatch-Emp} \citep{Yee2017}, which constructs a best-matching template as a linear combination of similar stars and uses that combination to infer the spectroscopic parameters.

We then inferred the fundamental stellar parameters with a grid-based modelling approach, by identifying the evolutionary models that best reproduce the observed spectroscopic constraints. This inference was carried out with the BAyesian STellar Algorithm (\texttt{BASTA}; \citealt{Aguirre22}) using the stellar isochrones of \citet{Hidalgo18}. In addition to the spectroscopic constraints above, we included the parallax and the $G_\mathrm{BP}$, $G_\mathrm{RP}$, and $G$ magnitudes to constrain the stellar luminosity \citep{Aguirre22}. Here, we adopted a minimum uncertainty of $0.01$ mag for all magnitudes \citep{Riello21}.

\subsection{Spectroscopic vetting and classification}
\label{sec:vetting}

\begin{figure*}
\centering
\includegraphics[width=18cm]{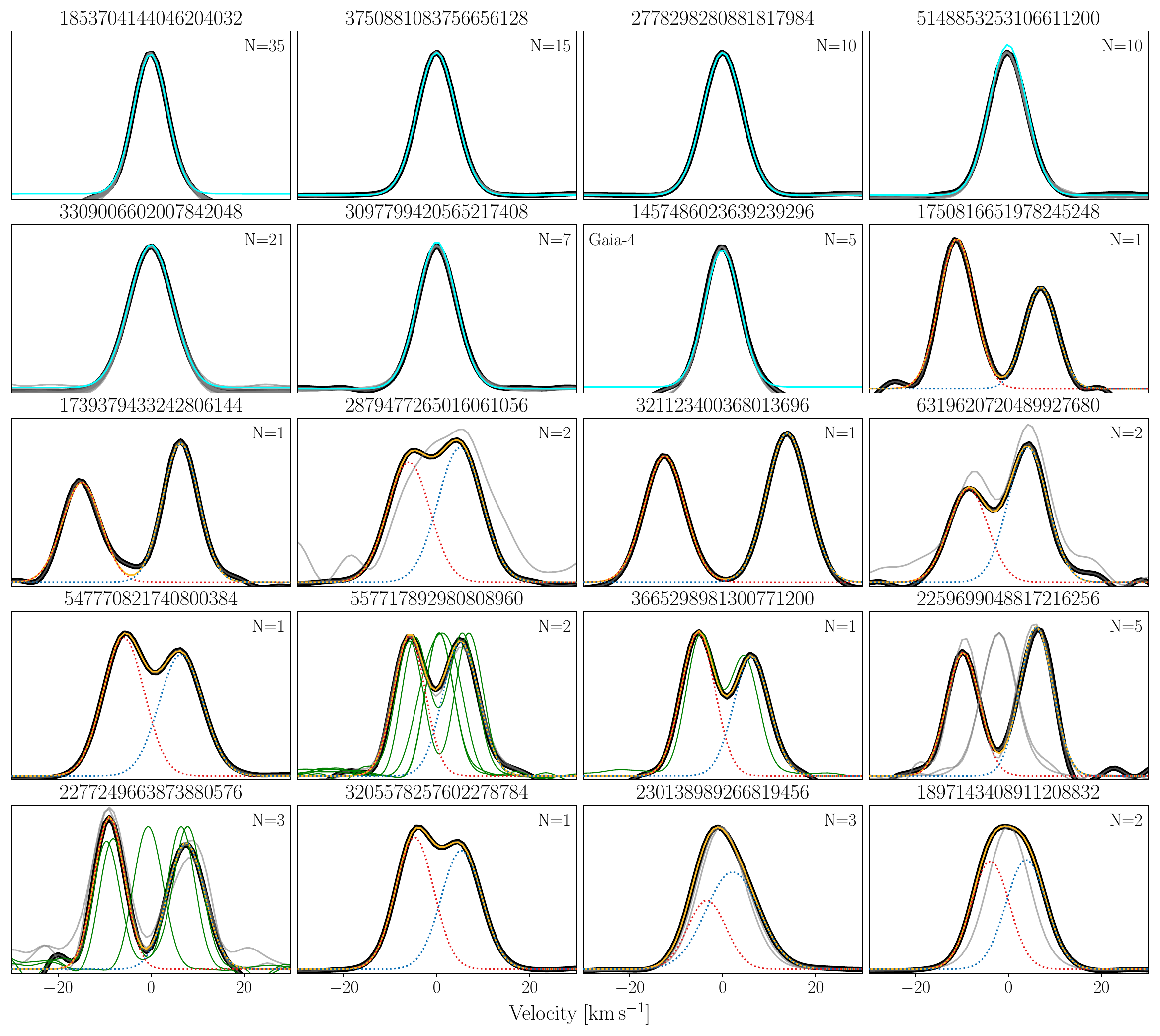}
\caption{Spectral broadening functions for the 20 observed systems. The total number of observations ($N$) is indicated in the top right corner of each panel. BFs for all epochs are superimposed; a representative observation is shown in solid black, with the remaining epochs in thin grey (NEID observations in green). The best-fit models are overlaid: cyan indicates a single-Gaussian model, while orange indicates a two-Gaussian model. For the latter, the individual Gaussian components are depicted as dotted lines in red and blue.}
\label{fig:bf_gallery}
\end{figure*}

We used broadening functions (BFs, \citealt{Rucinski1999}) to visualise the line profiles and identify binary systems. A BF is the convolution kernel that maps a template spectrum onto the observed spectrum. We computed the BFs using synthetic ATLAS9 spectra with spectroscopic parameters matching those derived for each system \citep{Kurucz1992}. Systems with two BF maxima separated by more than the width of the single-star BF peak were classified as binaries. When the velocity separation was smaller than this scale, the two components were not resolved as separate maxima. In these cases, we classified systems with gross asymmetries or time-variable changes in the BF morphology as binaries. Examples are shown in Fig.~\ref{fig:bf_gallery}.

\subsection{Orbital modelling}
\label{sec:forward}

Our orbital modelling follows the general approach of \citet{Winn2022}, \citet{Fitzmaurice2024}, and \citet{Stefansson2025}. Here we briefly outline the framework, summarise the adopted notation, and highlight the choices specific to this work.

Throughout, we considered three complementary fits: RV-only, astrometry-only, and joint RV + astrometry. The model parameters were
\begin{equation}
\Bigl\{
P,\:T_0,\:e,\:\omega,\:K,\:
i,\:\Omega,\:\varpi,\:
\vphantom{\sigma_\mathrm{jit}} \gamma,\:\sigma_\mathrm{jit}
\Bigr\},
\end{equation}
where $P$ is the orbital period, $T_0$ the time of periastron passage (measured from the \textit{Gaia} DR3 reference epoch, $\mathrm{J}2016.0$), $e$ the orbital eccentricity, $\omega$ the argument of periastron, $K$ the RV semi-amplitude, $i$ the orbital inclination, $\Omega$ the longitude of the ascending node, and $\varpi$ the system parallax. All orbital parameters refer to the motion of the host star. Nuisance parameters are the RV offset, $\gamma$, and the RV jitter, $\sigma_\mathrm{jit}$, which are defined in Sect.~\ref{sec:rvonly}. We fixed the parallax to the \textit{Gaia} DR3 value because its fractional uncertainty is negligible for our analysis. In the RV-only fit, we omitted the subset $(i,\Omega)$ as these parameters cannot be determined from RV data alone.

We adopted the \textit{Gaia} coordinate convention, using a right-handed coordinate system in which the $+Z$ axis points away from the observer, consistent with the conventions described by \citet{Householder2022} and \citet{Marcussen2026}. The ascending node is defined as the point where the orbit crosses $Z=0$ while the luminous body is moving away from the observer, and $\Omega$ is measured from north through east to that node.

In the joint analysis, we treated the RVs and astrometry as independent constraints on the same physical orbit and combined them by multiplying their likelihoods:
\begin{equation}
\mathcal{L}_\mathrm{tot} = \mathcal{L}_\mathrm{RV}\,\mathcal{L}_\mathrm{ast},
\end{equation}
where $\mathcal{L}_\mathrm{tot}$ is the total likelihood, and $\mathcal{L}_\mathrm{RV}$ and $\mathcal{L}_\mathrm{ast}$ quantify the agreement between the model and the radial velocity and astrometric data, respectively.

\subsubsection{RV model}
\label{sec:rvonly}
We modelled the RV time series assuming a single companion on a Keplerian orbit. The host-star radial velocity at time $t$ is
\begin{equation}
v(t) = \gamma + K\left[\cos\!\left(\nu+\omega\right) + e\cos\omega\right],
\label{eq:rvmodel}
\end{equation}
where $\nu(t)$ is the time-dependent true anomaly derived from $(P,T_0,e)$. To account for unmodelled instrumental and astrophysical noise, we added the RV jitter term, $\sigma_\mathrm{jit}$, in quadrature to the formal uncertainties, $\sigma_{v}$:
\begin{equation}
\sigma_{v,\mathrm{eff}}^2 = \sigma_{v}^2 + \sigma_\mathrm{jit}^2.
\end{equation}
These effective uncertainties defined the Gaussian likelihood used for the RV data.

\subsubsection{Astrometry}
\label{sec:gaialike}

Because \textit{Gaia} DR3 does not provide epoch astrometry, we could not construct the astrometric likelihood directly from time-series residuals, as we did for the RV data. Instead, following \citet{Winn2022}, we treated the published orbital solution and its covariance matrix as the astrometric constraint, assuming Gaussian errors. The covariance matrix can be reconstructed from the correlation matrix (using \texttt{nsstools}\footnote{\url{https://www.cosmos.esa.int/web/gaia/dr3-nss-tools}}) published in the \texttt{gaiadr3.nss\_two\_body\_orbits} table (see \citealt{Winn2022} for details).

\textit{Gaia} DR3 reports orbital solutions in the Thiele--Innes coefficients $(A,\,B,\,F,\,G)$. We therefore write the published DR3 solution as
\begin{equation}
\boldsymbol{\theta}_\mathrm{Gaia} = (A,\,B,\,F,\,G,\,P,\,T_0,\,e,\,\varpi).
\end{equation}
Because our inference framework is expressed in the Campbell elements, we mapped each sampled orbit into the corresponding Thiele--Innes representation. Specifically, for a given set of $(\omega,\Omega,i)$ and angular semi-major axis $a_1$ of the host star's orbit, we computed
\begin{align}
A &= a_1\left(\cos\Omega\cos\omega - \sin\Omega\sin\omega\cos i\right),\\
B &= a_1\left(\cos\Omega\sin\omega + \sin\Omega\cos\omega\cos i\right),\\
F &= a_1\left(-\sin\Omega\cos\omega - \cos\Omega\sin\omega\cos i\right),\\
G &= a_1\left(-\sin\Omega\sin\omega + \cos\Omega\cos\omega\cos i\right),
\end{align}
with
\begin{equation}
a_1 = \varpi\,\frac{P K}{2\pi}\frac{\sqrt{1-e^2}}{\sin i}.
\end{equation}

Strictly speaking, \textit{Gaia} solutions of the \texttt{astrometric} type describe the motion of the photocentre rather than the barycentric orbit of the primary star itself. In our modelling, however, we assumed that the photocentre coincided with the host star, meaning that the companion flux ratio is zero. This is the same dark-companion assumption used for the masses listed in Table~\ref{tab:parent_sample_overview}. For the systems modelled here, this is a reasonable approximation: their spectra show only one set of absorption lines, with no detectable contribution from a luminous secondary. In addition, the \textit{Gaia} and RV solutions are mutually consistent and imply substellar companion masses, for which the companion light is expected to be negligible relative to that of the host star.

The astrometric likelihood was then evaluated from the difference between $\boldsymbol{\theta}_\mathrm{Gaia}$ and the corresponding model vector, using the reconstructed covariance matrix. Evaluating the likelihood in this way allowed us to compare each sampled orbit directly to the catalogue-level astrometric solution reported by \textit{Gaia} DR3.

\subsubsection{Sampling and priors}
\label{sec:priors}

We assumed Gaussian observational uncertainties and sampled the posterior with the affine-invariant Markov chain Monte Carlo (MCMC) sampler \texttt{emcee} \citep{Foreman-Mackey2013}. Although the forward model is written in physical parameters, we sampled several transformed variables for efficiency and to impose weakly informative priors \citep[e.g.][]{Ford2006,Eastman2013}:

\begin{equation}
\{\log P,\;\log K,\;\sqrt{e}\cos\omega,\;\sqrt{e}\sin\omega,\;\cos i,\;\log\sigma_\mathrm{jit}\}.
\end{equation}
Sampling in \(\log P\), \(\log K\), and \(\log \sigma_\mathrm{jit}\) implies Jeffreys priors \citep{Jeffreys1946}, \(p(x)\propto 1/x\). Such log-flat priors are less informative than flat priors, which concentrate more of the prior weight near the upper bounds. Sampling uniformly in \(\cos i\) corresponds to an isotropic prior on orbital orientation, while the $(\sqrt{e}\cos\omega,\sqrt{e}\sin\omega)$ parametrisation improves sampling near \(e=0\). We initially treated the RV jitter, $\sigma_\mathrm{jit}$, as a free parameter in all fits. If the posterior was consistent with $\sigma_\mathrm{jit}=0$, we reran the fit with the jitter fixed to zero and adopted that solution in the final analysis.

\subsubsection{Derived quantities}
\label{sec:derived}

From the fitted parameters and the inferred host-star masses, we derived the companion masses. We first computed the binary mass function, $f(m)$:
\begin{equation}
f(m) \equiv \frac{m_2^3 \sin^3 i}{(m_1+m_2)^2}
= \frac{P K^3}{2\pi G}\left(1-e^2\right)^{3/2},
\end{equation}
which we then solved numerically for the companion mass, $m_2$, using Newton--Raphson iteration.

\section{Results}
\label{sec:results}

\subsection{Spectroscopic classifications and false positives}
\label{sec:falsepos}

Of the 20 observed systems, 13 show spectroscopic evidence of being near-equal-flux binaries and are therefore false positives in the search for substellar companions. Figure~\ref{fig:bf_gallery} summarises this false-positive sample and shows the corresponding broadening functions. In 11 of the 13 false positives, two distinct peaks are separated by more than one single-star BF peak width. For the remaining two (HD 275921 and HD 207740), however, line profiles are asymmetric and phase-dependent, indicative of spectral blending; these two stars appear in the final two panels of Fig.~\ref{fig:bf_gallery}. We also compared the observed velocity separations of the binary components with those predicted by the \textit{Gaia} astrometric solutions, assuming a mass ratio of unity. The measured separations are broadly consistent with the predictions. However, the propagated uncertainties are large because the uncertainties in $P$ and $T_0$ accumulate over the long interval since the \textit{Gaia} reference epoch. The DR3 solutions therefore provided little predictive power for identifying conjunctions or scheduling observations near maximum velocity separation. This may improve with \textit{Gaia} DR4.\footnote{\url{https://www.cosmos.esa.int/web/gaia/data-release-4}}

\subsection{Seven confirmed substellar companions}
\label{sec:confirmed}

\begin{figure*}
\centering
\includegraphics[width=18cm]{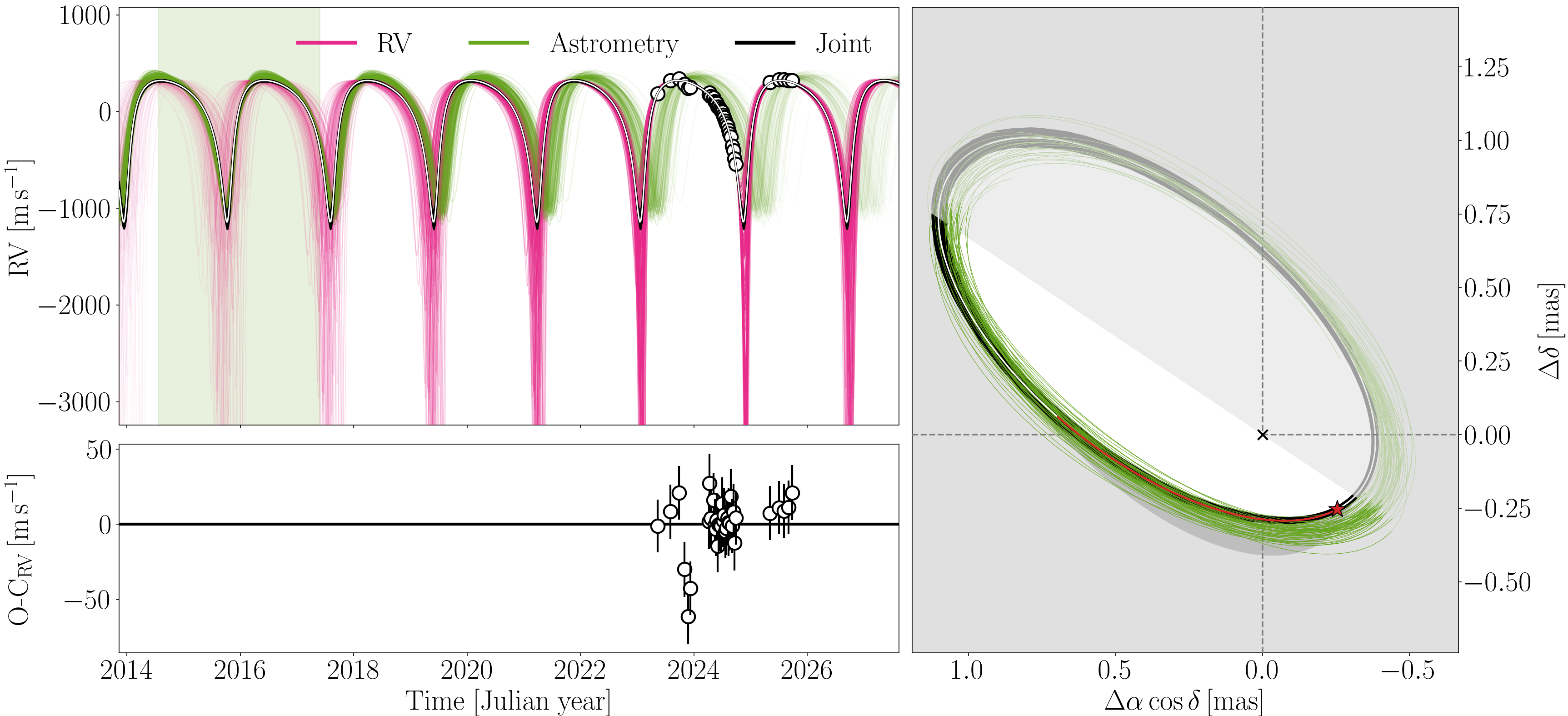}
\caption{Radial velocity data and model fit for LP 341-28. Top left: Radial velocity time series. Observations are shown as white points with black outlines, 150 random draws from the astrometry-only posterior distribution are shown in green, and RV-only draws are shown in pink. The median joint-fit model is shown in white, and the black shaded region indicates the $1\sigma$ uncertainty. The green-shaded interval marks the \textit{Gaia} DR3 observation window. Bottom left: Residuals relative to the joint fit. The displayed data-point uncertainties include jitter terms. Right: Sky-plane geometry of the host-star orbit. The median joint-fit solution is shown in white, with the black shaded band indicating the $1\sigma$ uncertainty. Fifty random draws from the astrometry-only posterior are shown in green. The red star marks periastron, and the solid red segment traces the motion over the next three months. The white part of the orbit lies on our side of the sky plane, with the line of nodes marking the transition between white and grey. The black cross indicates the barycentre.}
\label{fig:rv_gaia_4032}
\end{figure*}

\begin{figure*}
\centering
\includegraphics[width=18cm]{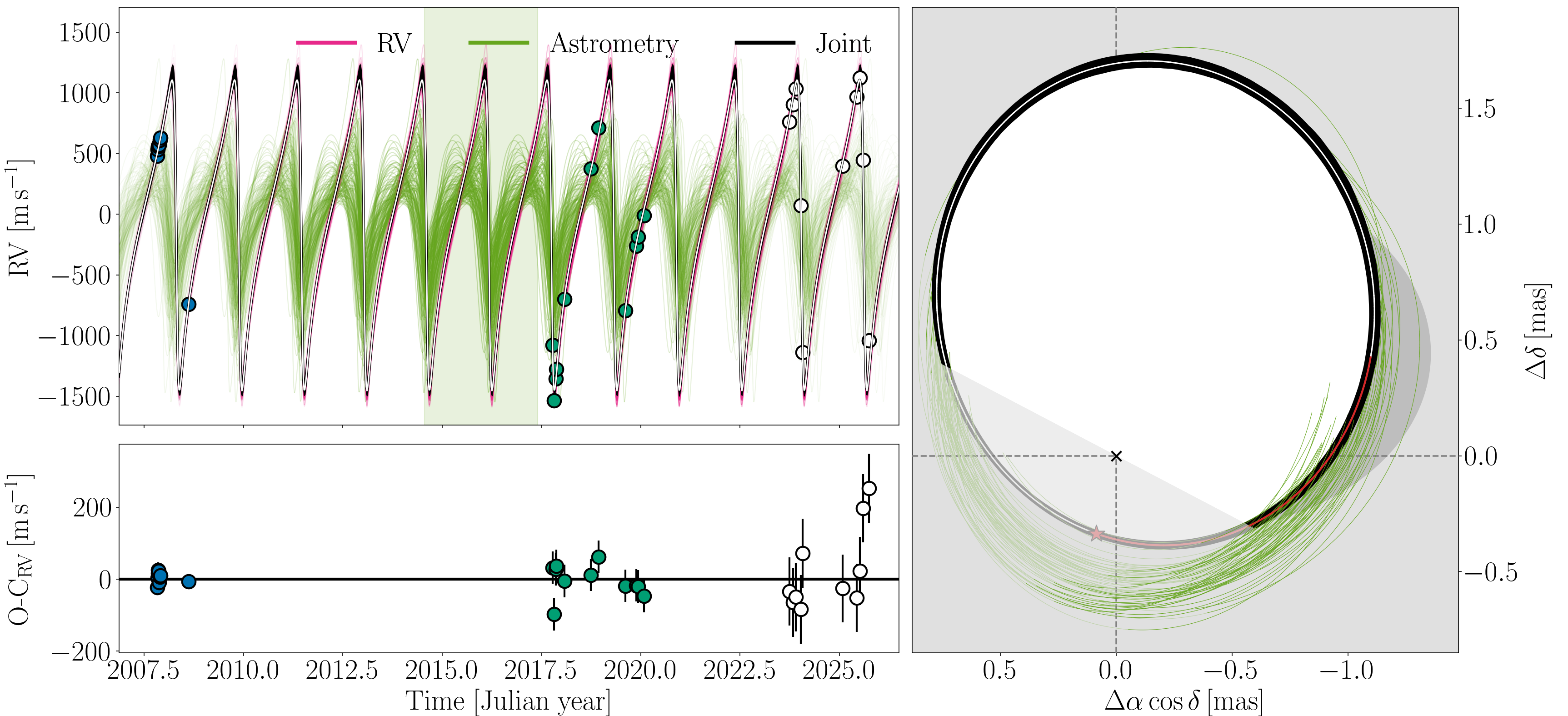}
\caption{Same as Fig.~\ref{fig:rv_gaia_4032} but for HD 5433 (Gaia DR3 2778298280881817984). Blue points show archival SOPHIE data and green points show archival SOPHIE+ data. White points show our FIES data.}
\label{fig:rv_gaia_7984}
\end{figure*}

\begin{figure*}
\centering
\includegraphics[width=18cm]{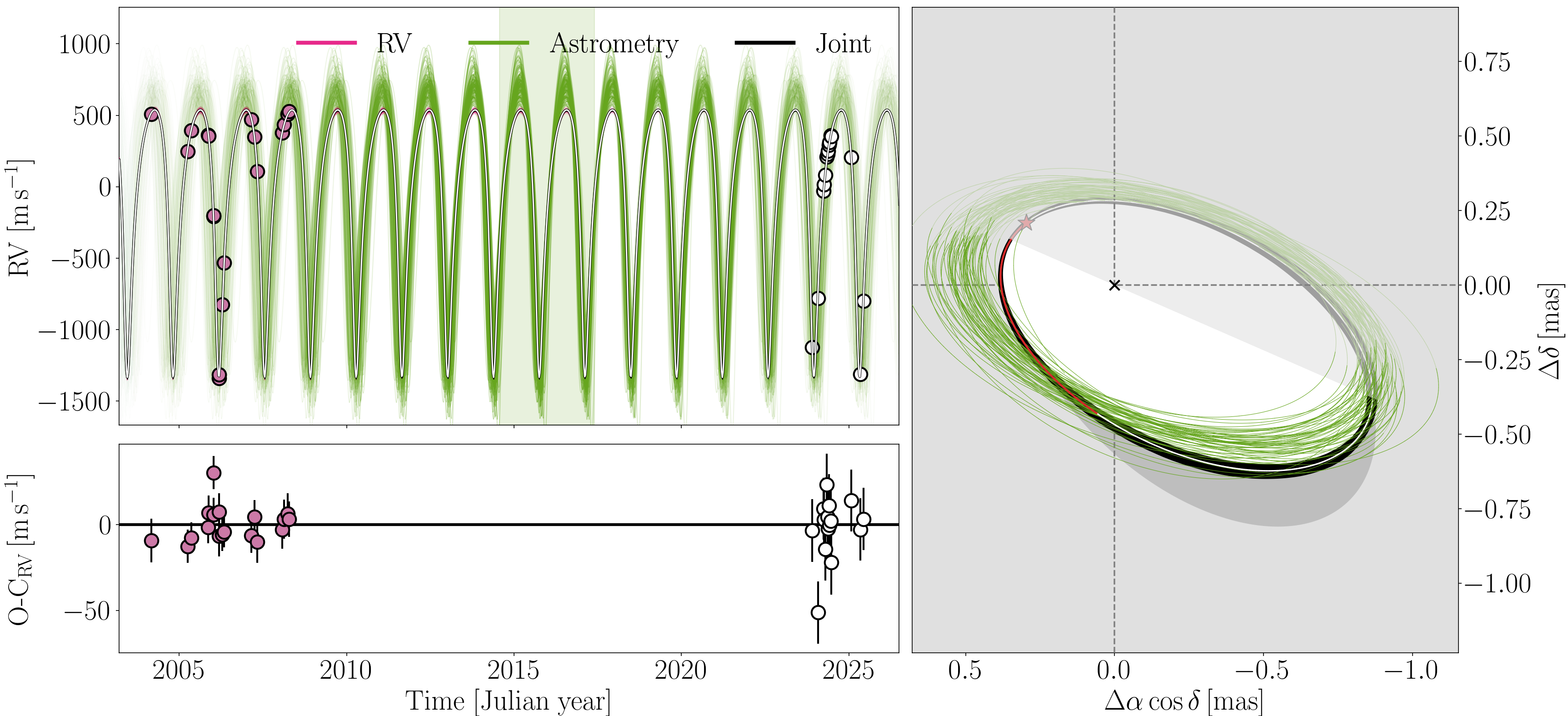}
\caption{Same as Fig.~\ref{fig:rv_gaia_4032} but for HD 91669 (Gaia DR3 3750881083756656128). Red points show Tull data, and white points show our FIES data.}
\label{fig:rv_gaia_6128}
\end{figure*}

\begin{figure*}
\centering
\includegraphics[width=18cm]{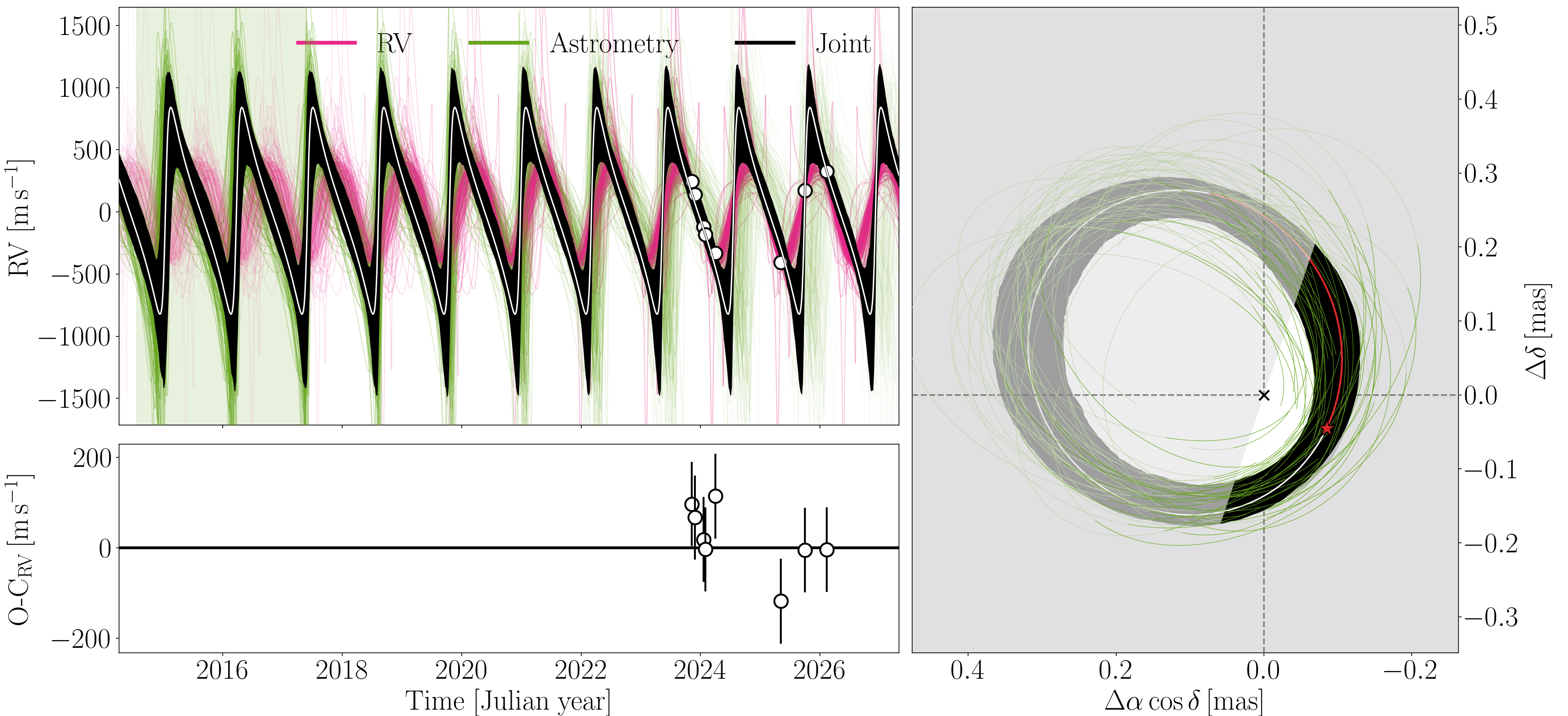}
\caption{Same as Fig.~\ref{fig:rv_gaia_4032} but for 2M0809+07 (Gaia DR3 3097799420565217408).}
\label{fig:rv_gaia_7408}
\end{figure*}

\begin{figure*}
\centering
\includegraphics[width=18cm]{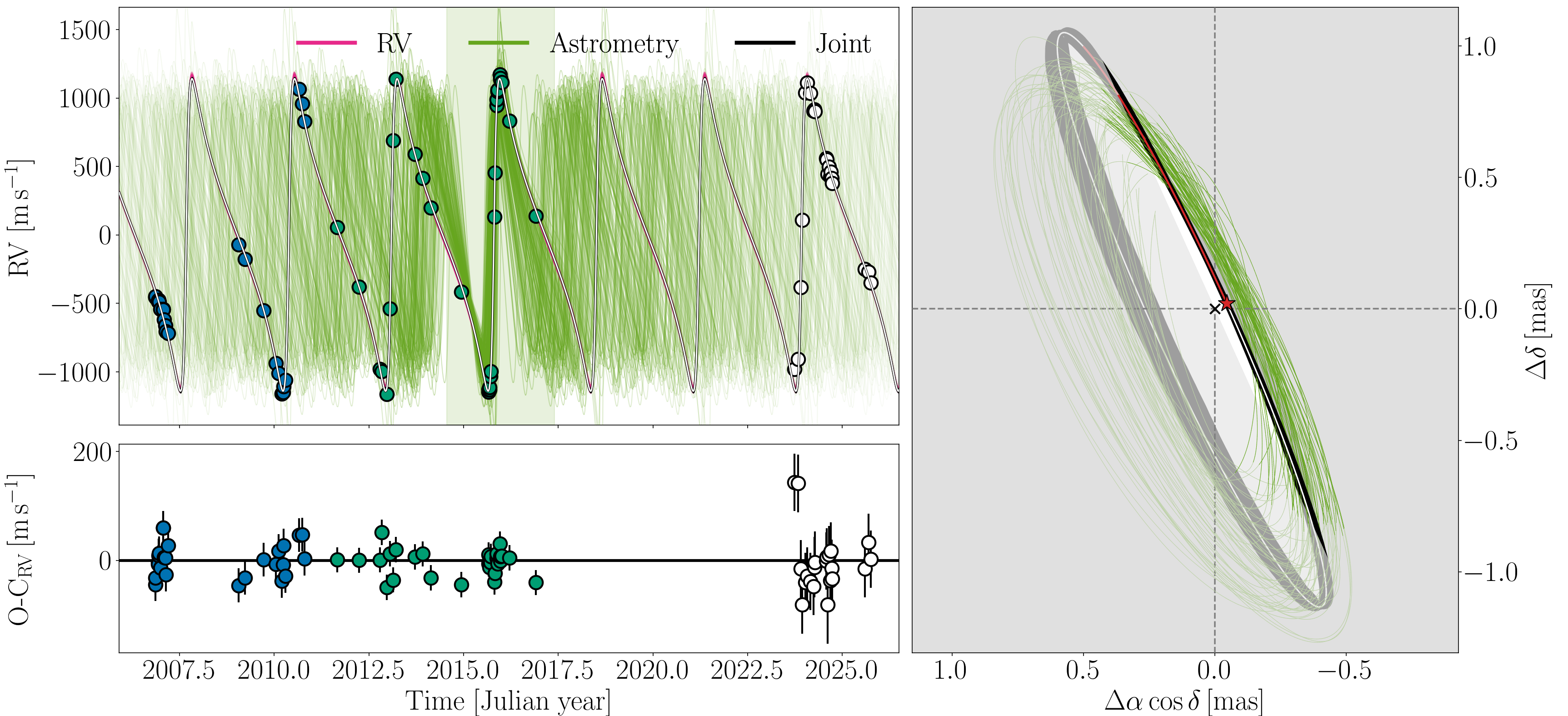}
\caption{Same as Fig.~\ref{fig:rv_gaia_4032} but for HD 30246 (Gaia DR3 3309006602007842048). Blue points show archival SOPHIE data and green points show archival SOPHIE+ data. White points show our FIES data.}
\label{fig:rv_gaia_2048}
\end{figure*}

\begin{figure*}
\centering
\includegraphics[width=18cm]{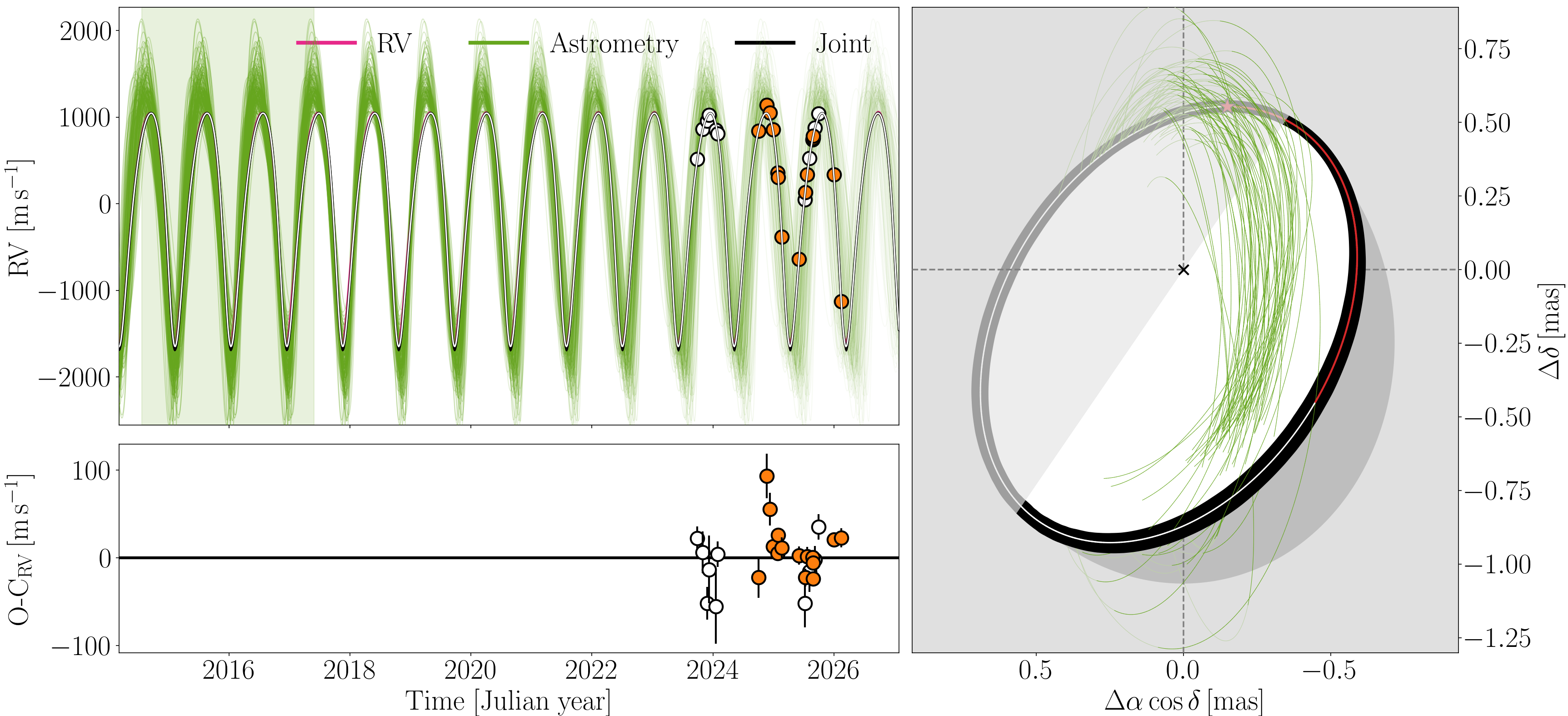}
\caption{Same as Fig.~\ref{fig:rv_gaia_4032} but for LP 769-9 (Gaia DR3 5148853253106611200). Orange points show the NIRPS data, and white points show the FIES data.}
\label{fig:rv_gaia_11200}
\end{figure*}

Seven systems remain consistent with a single-lined host star and show RV variations compatible with the \textit{Gaia} astrometric orbit. Below we describe six of them in detail. We do not discuss the seventh system, Gaia-4, because the FIES observations were already described by \citet{Stefansson2025}. For HD 91669, HD 5433, and HD 30246, we fitted the archival RVs from Tull and SOPHIE together with our own observations. For each system, we show (i) the RV-only solution, (ii) the RV curve predicted by the astrometry-only posterior, and (iii) the joint astrometry + RV solution. We also visualise the final three-dimensional orbital configuration from the joint fit alongside orbits drawn from the astrometry-only posterior. Figures~\ref{fig:rv_gaia_4032}--\ref{fig:rv_gaia_11200} show these comparisons.

Table~\ref{tab:finalparams_joint_all} summarises the final joint-fit parameters for all confirmed systems and the derived companion masses. Figures~\ref{fig:corner7984}--\ref{fig:corner7408} show corner plots from the MCMC sampling of all three fits and illustrate the overall consistency of the inferred parameters across the different data combinations. The joint fits substantially reduce the uncertainties in the orbital parameters shared with the astrometry-only fits. The median improvement was a factor of $18.0$ for $P$, $5.6$ for $T_0$, $10.0$ for $e$, and $7.7$ for $\omega$. The uncertainty in the RV semi-amplitude $K$ improved by a factor of $10.5$ relative to the value implied by the \textit{Gaia} astrometry. Notably, even the uncertainties of $\Omega$ and $i$ shrank substantially once the RV data tighten the shared orbital parameters, even though $\Omega$ and $i$ are constrained only by the astrometry. For example, the uncertainty in $i$ for the system HD 5433 decreases from $8\,\mathrm{deg}$ in the DR3 solution to $0.6\,\mathrm{deg}$ in the joint fit (Fig.\ref{fig:corner7984}). Across the sample, the median improvement in the uncertainties of $\Omega$ and $i$ was a factor of two.

\begin{table*}
\caption{Joint posterior medians and 68\% credible intervals for systems LP 341-28, HD 91669, HD 5433, LP 769-9, HD 30246, 2M0809+07.}
\label{tab:finalparams_joint_all}
\centering
\resizebox{\textwidth}{!}{%
\begin{tabular}{l c c c c c c}
\hline\hline
Parameter & LP 341-28 & HD 91669 & HD 5433 & LP 769-9 & HD 30246 & 2M0809+07 \\
\hline
$P$ [d] & $665.32 \pm 0.71$ & $499.250 \pm 0.050$ & $574.63 \pm 0.41$ & $337.75 \pm 0.24$ & $989.20 \pm 0.20$ & $433.8 \pm 2.4$ \\
$t_{\mathrm{p}}$ [BJD$ - 2457389$] & $-77.9 \pm 3.3$ & $-98.10 \pm 0.60$ & $66.2^{+1.6}_{-2.0}$ & $-3.2 \pm 2.7$ & $-72.30 \pm 0.40$ & $83 \pm 12$ \\
$e$ & $0.586 \pm 0.015$ & $0.4490 \pm 0.0030$ & $0.662 \pm 0.014$ & $0.2520 \pm 0.0040$ & $0.6630 \pm 0.0030$ & $0.56 \pm 0.12$ \\
$\omega$ [deg] & $198.1 \pm 1.0$ & $161.80 \pm 0.60$ & $100.5 \pm 2.3$ & $153.8 \pm 1.0$ & $270.1 \pm 0.5$ & $262 \pm 21$ \\
$K$ [m\,s$^{-1}$] & $727 \pm 34$ & $934.0 \pm 4.0$ & $1287 \pm 37$ & $1355 \pm 11$ & $1141.0 \pm 5.0$ & $800^{+250}_{-190}$ \\
$\sigma_{\mathrm{jit}}$ [m\,s$^{-1}$] & $17.3^{+2.8}_{-2.2}$ & $17.8^{+4.3}_{-3.1}$ & $94^{+22}_{-17}$ & ... & $52.6^{+9.0}_{-7.6}$ & ... \\
$i$ [deg] & $127.3 \pm 1.9$ & $51.9 \pm 1.2$ & $38.30 \pm 0.60$ & $135.8 \pm 1.6$ & $84.7 \pm 1.2$ & $33.5 \pm 5.9$ \\
$\Omega$ [deg] & $56.3 \pm 1.7$ & $246.4 \pm 1.6$ & $62.3 \pm 2.3$ & $145.6 \pm 2.1$ & $24.4 \pm 1.4$ & $341 \pm 25$ \\
\hline
\multicolumn{7}{l}{\textit{Derived}} \\
$a_{1}$ [mas] & $0.895 \pm 0.017$ & $0.679 \pm 0.011$ & $1.297 \pm 0.016$ & $0.806 \pm 0.022$ & $1.6000 \pm 0.0090$ & $0.259 \pm 0.021$ \\
$a$ [mas] & $26.50^{+0.20}_{-0.20}$ & $17.30^{+0.10}_{-0.20}$ & $22.00 \pm 0.20$ & $11.90 \pm 0.10$ & $40.40^{+0.50}_{-0.70}$ & $5.70 \pm 0.10$ \\
$a/\varpi$ [$\mathrm{au}$] & $1.339^{+0.010}_{-0.012}$ & $1.241^{+0.010}_{-0.016}$ & $1.397 \pm 0.015$ & $0.8610 \pm 0.0070$ & $1.972^{+0.027}_{-0.032}$ & $1.049 \pm 0.016$ \\
$f(m)$ [$10^{-6}\,M_\odot$] & $14.1 \pm 1.6$ & $30.10 \pm 0.40$ & $53.6 \pm 2.5$ & $78.9 \pm 1.8$ & $63.9 \pm 1.1$ & $13.0^{+9.6}_{-6.1}$ \\
$m_2$ [$M_\mathrm{Jup}$] & $25.60 \pm 0.60$ & $42.0 \pm 1.1$ & $67.9 \pm 1.7$ & $53.0 \pm 1.7$ & $43.3 \pm 1.3$ & $39.0 \pm 3.5$ \\
\hline
\end{tabular}%
}
\end{table*}

\section{Discussion}
\label{sec:discussion}

\subsection{Orbital solution robustness}

\begin{figure}
\centering
\includegraphics[width=\columnwidth]{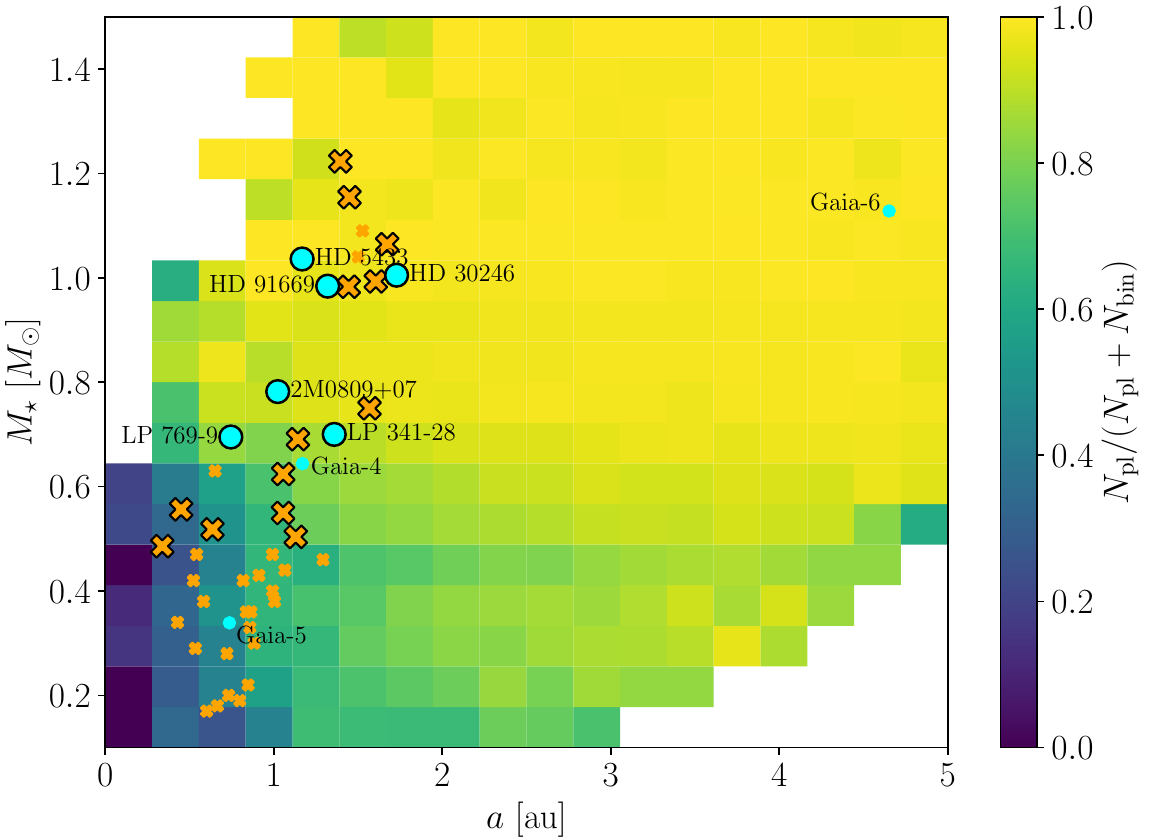}
\caption{Expected genuine-companion fraction and observed classifications. Our six confirmed brown dwarfs are shown as large cyan circles with black outlines, and false positives as large orange crosses with black outlines. Smaller symbols in the same colour scheme without outlines indicate confirmed companions and false positives from \citet{Stefansson2025} and \citet{Pinamonti2026}. The symbols are overplotted on a map of the expected \textit{Gaia} DR5 genuine-planet fraction recreated from \citet{Lammers2026}.}
\label{fig:fp}
\end{figure}

Joint analyses of RVs and \textit{Gaia} DR3 orbital solutions assume that the published covariance matrix gives a realistic description of the parameter uncertainties. Because that assumption was called into question by some earlier studies (e.g., \citealt{Winn2022,Sozzetti2023,Fitzmaurice2024,Pinamonti2026}), we initially allowed for an additional covariance-inflation parameter following \citet{Fitzmaurice2024}. However, for all systems in our final sample, the posterior for this parameter was consistent with unity, so we ultimately adopted the nominal \textit{Gaia} covariance matrices. A likely reason for the higher degree of consistency is that our confirmed systems lie towards the high-mass, high-significance end of the DR3 substellar-candidate sample. In this regime, the astrometric signal is larger relative to the \textit{Gaia} uncertainties than for the lower-mass candidates examined in earlier follow-up studies. The corresponding orbital solutions may therefore be less vulnerable to degeneracies and other failure modes.

\subsection{False positive rate}
\citet{Lammers2026} simulated the expected planet yield and false-positive fraction for \textit{Gaia} DR4 and DR5, finding that the false positive rate
should be greatest for candidates with relatively short-period orbits and low-mass host stars. Astrometry is inherently more sensitive to wider orbits because these produce larger reflex signals on the sky -- as long as the time baseline of observations covers at least one full orbit. Because the DR3 baseline was only 33 months, the period range of the substellar candidates is limited to the regime with a high false-positive rate. The longer baselines of DR4 and DR5 should make it possible to detect wider-orbit systems, and the overall false-positive rate is expected to decrease. Figure~\ref{fig:fp} reproduces Figure~12 from \citet{Lammers2026}, showing the fraction of genuine planet and brown-dwarf detections across this parameter space, and overplots both the systems from \citet{Stefansson2025} and the 20 systems analysed in this work. Our results are in agreement
with the predicted trend.

An instructive individual case is HD\,68638. Based on double-peaked CCFs in archival ELODIE spectra, \citet{Holl2023} and \citet{MarcussenAlbrecht2023} identified this \textit{Gaia} DR3 candidate as a double-lined binary and hence an astrometric false positive. \citet{Unger2023}, however, modelled RVs derived from the same ELODIE spectra jointly with the DR3 orbital solution, incorrectly concluding that the companion is a ${\sim}35\,M_\mathrm{Jup}$ brown dwarf.
This shows that apparent agreement between the RV and astrometric signals does not necessarily provide an independent validation. This is because, in a near-twin binary, the photocentre motion measured by \textit{Gaia} and the RV signal inferred by fitting a single profile to the blended spectra are diluted to nearly the same degree \citep{MarcussenAlbrecht2023}. The two spurious signals can therefore mimic a consistent substellar orbit, allowing a joint fit to converge on a well-behaved but incorrect solution. Agreement between RVs and astrometry is thus not sufficient on its own to exclude this type of false positive, and inspection of the line profiles remains necessary.

\subsection{Brown dwarf desert}

\begin{figure}
\centering
\includegraphics[width=\columnwidth]{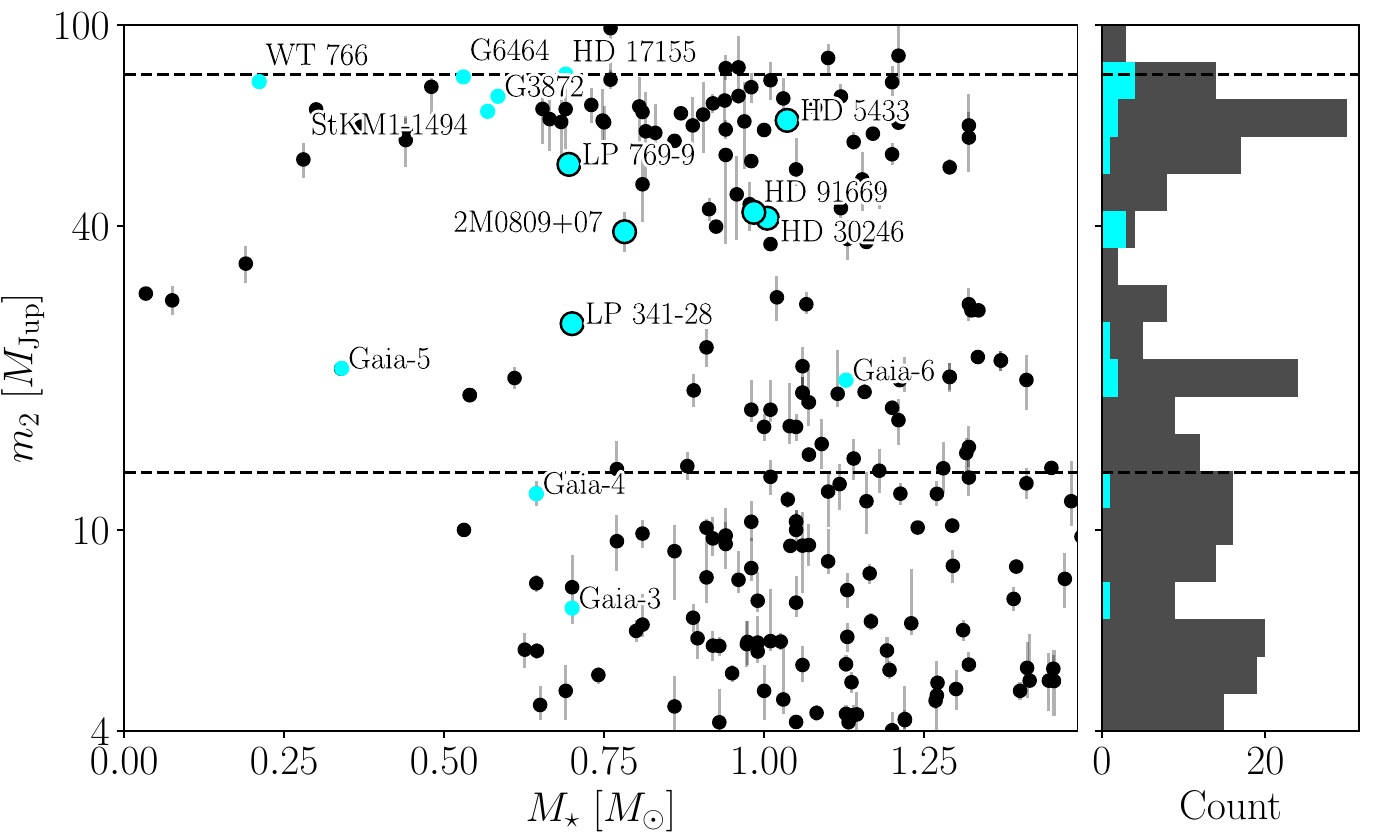}
\caption{Companion mass as a function of host-star mass for known exoplanets and brown dwarfs. Exoplanet data are from the NASA Exoplanet Archive \citep{Akeson2013} and brown dwarfs are from \citet{Stevenson2023} plus five from \citet{Winterhalder2024}. Minimum effective masses, $m_2\sin i$, are not included. Our six confirmed brown dwarfs are shown as large cyan circles, while previously confirmed \textit{Gaia} DR3 astrometric exoplanets and brown dwarfs are shown as smaller cyan circles.}
\label{fig:desert}
\end{figure}

To place our confirmed companions in a broader demographic context, we compile known exoplanets and brown dwarfs and display them in a companion--host mass diagram (Fig.~\ref{fig:desert}). In this representation, the `brown-dwarf desert' \citep{Marcy2000,Grether2006,2014Ma,Stevenson2023} appears as a sparsely populated region at intermediate companion masses, between the abundant giant-planet population and the low-mass stellar-companion population. Of our six systems, several fall squarely within this underpopulated regime, with true companion masses of $\sim$26--68~$M_{\mathrm{Jup}}$ around roughly solar-mass hosts. Our confirmed sample therefore probes a region where multiple formation channels may overlap, spanning the high-mass tail of planet formation (e.g.\ core accretion) and the low-mass end of stellar-like formation (e.g.\ disc gravitational instability or fragmentation).

\section{Summary}
We carried out an RV follow-up campaign of 20 \textit{Gaia} astrometric substellar candidates using FIES, NIRPS and NEID. Spectroscopic vetting proved essential: 13 systems were identified as stellar binaries, showing that unresolved binaries remain a major source of false positives among \textit{Gaia} DR3 substellar candidates. Of the systems that survived this step, seven remained consistent with single-lined spectra. For one of these seven candidates, Gaia-4, the exoplanet companion had already been confirmed by \cite{Stefansson2025}. 
For the remaining six systems, we jointly model the RVs and \textit{Gaia} astrometry. For HD 5433, HD 91669, and HD~30246, this confirms the brown dwarf classifications reported by \citet{Unger2023} and refines their orbital parameters. For LP 341-28, recently confirmed by \citet{Barbato2026}, we obtain consistent results. For LP 769-9 and 2M0809+07, we report the first confirmations of their brown dwarf companions.
The companion masses are between 26 and 68 Jupiter masses.

For these confirmed systems, the \textit{Gaia} DR3 orbital solutions and their published uncertainties are broadly consistent with the RV data, in contrast to the tensions reported for some lower-significance candidates in earlier studies. The high degree of consistency demonstrates that the high-significance end of the DR3 sample already provides a robust basis for orbit inference. Future \textit{Gaia} data releases, which will provide epoch astrometry, will make such analyses substantially more powerful by enabling fully unified fits to astrometry and RVs.

More broadly, our results fit well into the emerging picture of \textit{Gaia}'s substellar population. The high false-positive rate in DR3 is consistent with expectations for this early release, while future releases should improve as their longer baselines shift sensitivity towards wider orbits. At the same time, our confirmed companions in the brown-dwarf desert show that \textit{Gaia} is already reaching into sparsely explored regions of substellar parameter space.

\begin{acknowledgements}
We thank the anonymous referee for constructive comments that improved the manuscript. We thank the NEID Queue Observers and WIYN Observing Associates for their skillful execution of our observations. Data presented were obtained by the NEID spectrograph built by Penn State University and operated at the WIYN Observatory by NOIRLab, under the NN-EXPLORE partnership of the National Aeronautics and Space Administration and the National Science Foundation. The NEID archive is operated by the NASA Exoplanet Science Institute at the California Institute of Technology. Based in part on observations at the Kitt Peak National Observatory (Prop. ID 2023B-982758 and 2024B-726751), managed by the Association of Universities for Research in Astronomy (AURA) under a cooperative agreement with the National Science Foundation. The WIYN Observatory is a joint facility of the NSF's National Optical-Infrared Astronomy Research Laboratory, Indiana University, the University of Wisconsin-Madison, Pennsylvania State University, Purdue University, and Princeton University. The authors are honored to be permitted to conduct astronomical research on Iolkam Du'ag (Kitt Peak), a mountain with particular significance to the Tohono O'odham. Data presented herein were obtained from telescope time allocated to NN-EXPLORE through the scientific partnership of the National Aeronautics and Space Administration, the National Science Foundation, and the National Optical Astronomy Observatory.

Based on observations made with the Nordic Optical Telescope, owned in collaboration by the University of Turku and Aarhus University, and operated jointly by Aarhus University, the University of Turku and the University of Oslo, representing Denmark, Finland and Norway, the University of Iceland and Stockholm University at the Observatorio del Roque de los Muchachos, La Palma, Spain, of the Instituto de Astrofísica de Canarias. The NOT data were obtained under programs 67-008, 68-010, 69-010 and 71-018.

Based on observations collected at the European Southern Observatory under ESO programmes 114.2758, 115.27ZE, 116.296E, 117.2AJC.
The WIYN Observatory is a joint facility of the NSF’s National Optical Infrared Astronomy Research Laboratory, Indiana University, the University of Wisconsin–Madison, Pennsylvania State University and Princeton University.
This work has made use of data from the European Space Agency (ESA) mission \textit{Gaia} (\url{https://www.cosmos.esa.int/gaia}), processed by the \textit{Gaia} Data Processing and Analysis Consortium (DPAC, \url{https://www.cosmos.esa.int/web/gaia/dpac/consortium}). Funding for the DPAC has been provided by national institutions, in particular the institutions participating in the Gaia Multilateral Agreement.
\end{acknowledgements}

\bibliographystyle{aa}
\bibliography{biblio}

@ARTICLE{Barbato2026,
       author = {{Barbato}, D. and {Pinamonti}, M. and {Sozzetti}, A. and {Desidera}, S. and {D'Orazi}, V. and {Maldonado}, J. and {Biazzo}, K. and {Naponiello}, L. and {Lanza}, A.~F. and {Bignamini}, A. and {Bonomo}, A.~S. and {Brogi}, M. and {Cabona}, L. and {Damasso}, M. and {Gratton}, R. and {Mancini}, L. and {Mantovan}, G. and {Nardiello}, D. and {Rainer}, M. and {Guilluy}, G. and {Giacobbe}, P. and {Malavolta}, L. and {Cosentino}, R. and {Boschin}, W. and {Claudi}, R.},
        title = "{The GAPS programme at TNG: LXXV. Validating and confirming Gaia sub-stellar astrometric candidates with HARPS-N}",
      journal = {\aap},
         year = 2026,
        month = aug,
       volume = {712},
          eid = {A213},
        pages = {A213},
          doi = {10.1051/0004-6361/202557585},
archivePrefix = {arXiv},
       eprint = {2606.29001},
 primaryClass = {astro-ph.EP},
       adsurl = {https://ui.adsabs.harvard.edu/abs/2026A&A...712A.213B}
}

@ARTICLE{Winterhalder2024,
       author = {{Winterhalder}, T.~O. and {Lacour}, S. and {M{\'e}rand}, A. and {Kammerer}, J. and {Maire}, A.-L. and {Stolker}, T. and {Pourr{\'e}}, N. and {Babusiaux}, C. and {Glindemann}, A. and {Abuter}, R. and {Amorim}, A. and {Asensio-Torres}, R. and {Balmer}, W.~O. and {Benisty}, M. and {Berger}, J.-P. and {Beust}, H. and {Blunt}, S. and {Boccaletti}, A. and {Bonnefoy}, M. and {Bonnet}, H. and {Bordoni}, M.~S. and {Bourdarot}, G. and {Brandner}, W. and {Cantalloube}, F. and {Caselli}, P. and {Charnay}, B. and {Chauvin}, G. and {Chavez}, A. and {Choquet}, E. and {Christiaens}, V. and {Cl{\'e}net}, Y. and {Coud{\'e} du Foresto}, V. and {Cridland}, A. and {Davies}, R. and {Dembet}, R. and {Dexter}, J. and {Drescher}, A. and {Duvert}, G. and {Eckart}, A. and {Eisenhauer}, F. and {F{\"o}rster Schreiber}, N.~M. and {Garcia}, P. and {Garcia Lopez}, R. and {Gardner}, T. and {Gendron}, E. and {Genzel}, R. and {Gillessen}, S. and {Girard}, J.~H. and {Grant}, S. and {Haubois}, X. and {Hei{\ss}el}, G. and {Henning}, Th. and {Hinkley}, S. and {Hippler}, S. and {Houll{\'e}}, M. and {Hubert}, Z. and {Jocou}, L. and {Keppler}, M. and {Kervella}, P. and {Kreidberg}, L. and {Kurtovic}, N.~T. and {Lagrange}, A.-M. and {Lapeyr{\`e}re}, V. and {Le Bouquin}, J.-B. and {Lutz}, D. and {Mang}, F. and {Marleau}, G.-D. and {Molli{\`e}re}, P. and {Monnier}, J.~D. and {Mordasini}, C. and {Mouillet}, D. and {Nasedkin}, E. and {Nowak}, M. and {Ott}, T. and {Otten}, G.~P.~P.~L. and {Paladini}, C. and {Paumard}, T. and {Perraut}, K. and {Perrin}, G. and {Pfuhl}, O. and {Pueyo}, L. and {Ribeiro}, D.~C. and {Rickman}, E. and {Rustamkulov}, Z. and {Shangguan}, J. and {Shimizu}, T. and {Sing}, D. and {Stadler}, J. and {Straub}, O. and {Straubmeier}, C. and {Sturm}, E. and {Tacconi}, L.~J. and {van Dishoeck}, E.~F. and {Vigan}, A. and {Vincent}, F. and {von Fellenberg}, S.~D. and {Wang}, J.~J. and {Widmann}, F. and {Woillez}, J. and {Yazici}, {\c{S}}.},
        title = "{Combining Gaia and GRAVITY: Characterising five new directly detected substellar companions}",
      journal = {\aap},
         year = 2024,
        month = aug,
       volume = {688},
          eid = {A44},
        pages = {A44},
          doi = {10.1051/0004-6361/202450018},
archivePrefix = {arXiv},
       eprint = {2403.13055},
 primaryClass = {astro-ph.EP},
       adsurl = {https://ui.adsabs.harvard.edu/abs/2024A&A...688A..44W}
}

@ARTICLE{Unger2023,
       author = {{Unger}, N. and {S{\'e}gransan}, D. and {Barbato}, D. and {Delisle}, J.-B. and {Sahlmann}, J. and {Holl}, B. and {Udry}, S.},
        title = "{Exploring the brown dwarf desert with precision radial velocities and Gaia DR3 astrometric orbits}",
      journal = {\aap},
         year = 2023,
        month = dec,
       volume = {680},
          eid = {A16},
        pages = {A16},
          doi = {10.1051/0004-6361/202347578},
archivePrefix = {arXiv},
       eprint = {2310.02758},
 primaryClass = {astro-ph.EP},
       adsurl = {https://ui.adsabs.harvard.edu/abs/2023A&A...680A..16U}
}

@ARTICLE{Diaz2012,
       author = {{D{\'\i}az}, R.~F. and {Santerne}, A. and {Sahlmann}, J. and {H{\'e}brard}, G. and {Eggenberger}, A. and {Santos}, N.~C. and {Moutou}, C. and {Arnold}, L. and {Boisse}, I. and {Bonfils}, X. and {Bouchy}, F. and {Delfosse}, X. and {Desort}, M. and {Ehrenreich}, D. and {Forveille}, T. and {Lagrange}, A.-M. and {Lovis}, C. and {Pepe}, F. and {Perrier}, C. and {Queloz}, D. and {S{\'e}gransan}, D. and {Udry}, S. and {Vidal-Madjar}, A.},
        title = "{The SOPHIE search for northern extrasolar planets. IV. Massive companions in the planet-brown dwarf boundary}",
      journal = {\aap},
         year = 2012,
        month = feb,
       volume = {538},
          eid = {A113},
        pages = {A113},
          doi = {10.1051/0004-6361/201117935},
archivePrefix = {arXiv},
       eprint = {1111.1168},
 primaryClass = {astro-ph.EP},
       adsurl = {https://ui.adsabs.harvard.edu/abs/2012A&A...538A.113D}
}

@ARTICLE{Dalal2021,
       author = {{Dalal}, S. and {Kiefer}, F. and {H{\'e}brard}, G. and {Sahlmann}, J. and {Sousa}, S.~G. and {Forveille}, T. and {Delfosse}, X. and {Arnold}, L. and {Astudillo-Defru}, N. and {Bonfils}, X. and {Boisse}, I. and {Bouchy}, F. and {Bourrier}, V. and {Brugger}, B. and {Cort{\'e}s-Zuleta}, P. and {Deleuil}, M. and {Demangeon}, O.~D.~S. and {D{\'\i}az}, R.~F. and {Hara}, N.~C. and {Heidari}, N. and {Hobson}, M.~J. and {Lopez}, T. and {Lovis}, C. and {Martioli}, E. and {Mignon}, L. and {Mousis}, O. and {Moutou}, C. and {Rey}, J. and {Santerne}, A. and {Santos}, N.~C. and {S{\'e}gransan}, D. and {Str{\o}m}, P.~A. and {Udry}, S.},
        title = "{The SOPHIE search for northern extrasolar planets. XVII. A wealth of new objects: Six cool Jupiters, three brown dwarfs, and 16 low-mass binary stars}",
      journal = {\aap},
         year = 2021,
        month = jul,
       volume = {651},
          eid = {A11},
        pages = {A11},
          doi = {10.1051/0004-6361/202140712},
archivePrefix = {arXiv},
       eprint = {2105.09741},
 primaryClass = {astro-ph.EP},
       adsurl = {https://ui.adsabs.harvard.edu/abs/2021A&A...651A..11D}
}

@ARTICLE{GonzalezPayo2024,
  author = {{Gonz{\'a}lez-Payo}, J. and {Caballero}, J.~A. and {Gorgas}, J. and
            {Cort{\'e}s-Contreras}, M. and {G{\'a}lvez-Ortiz}, M.-C. and {Cifuentes}, C.},
  title = {{Multiplicity of stars with planets in the solar neighbourhood}},
  journal = {\aap},
  year = 2024,
  volume = {689},
  pages = {A302},
  doi = {10.1051/0004-6361/202450048}
}

@ARTICLE{Mason2001,
  author = {{Mason}, B.~D. and {Wycoff}, G.~L. and {Hartkopf}, W.~I. and {Douglass}, G.~G.
            and {Worley}, C.~E.},
  title = {{The 2001 US Naval Observatory Double Star CD-ROM. I. The Washington Double
            Star Catalog}},
  journal = {\aj},
  year = 2001,
  volume = {122},
  pages = {3466},
  doi = {10.1086/323920}
}

@ARTICLE{Bailer2026,
       author = {{Bailer-Jones}, C.~A.~L. and {Kreidberg}, L.},
        title = "{Component masses in stellar and sub-stellar binaries from Gaia astrometry and photometry}",
      journal = {\aap},
         year = 2026,
        month = apr,
       volume = {708},
          eid = {A249},
        pages = {A249},
          doi = {10.1051/0004-6361/202659004},
archivePrefix = {arXiv},
       eprint = {2603.03036},
 primaryClass = {astro-ph.SR},
       adsurl = {https://ui.adsabs.harvard.edu/abs/2026A&A...708A.249B}
}

@ARTICLE{Eastman2013,
       author = {{Eastman}, Jason and {Gaudi}, B. Scott and {Agol}, Eric},
        title = "{EXOFAST: A Fast Exoplanetary Fitting Suite in IDL}",
      journal = {\pasp},
         year = 2013,
        month = jan,
       volume = {125},
       number = {923},
        pages = {83},
          doi = {10.1086/669497},
archivePrefix = {arXiv},
       eprint = {1206.5798},
 primaryClass = {astro-ph.IM},
       adsurl = {https://ui.adsabs.harvard.edu/abs/2013PASP..125...83E}
}

@ARTICLE{Ford2006,
       author = {{Ford}, Eric B.},
        title = "{Improving the Efficiency of Markov Chain Monte Carlo for Analyzing the Orbits of Extrasolar Planets}",
      journal = {\apj},
         year = 2006,
        month = may,
       volume = {642},
       number = {1},
        pages = {505-522},
          doi = {10.1086/500802},
archivePrefix = {arXiv},
       eprint = {astro-ph/0512634},
 primaryClass = {astro-ph},
       adsurl = {https://ui.adsabs.harvard.edu/abs/2006ApJ...642..505F}
}

@ARTICLE{Chabrier2023,
       author = {{Chabrier}, Gilles and {Baraffe}, Isabelle and {Phillips}, Mark and {Debras}, Florian},
        title = "{Impact of a new H/He equation of state on the evolution of massive brown dwarfs. New determination of the hydrogen burning limit}",
      journal = {\aap},
         year = 2023,
        month = mar,
       volume = {671},
          eid = {A119},
        pages = {A119},
          doi = {10.1051/0004-6361/202243832},
archivePrefix = {arXiv},
       eprint = {2212.07153},
 primaryClass = {astro-ph.SR},
       adsurl = {https://ui.adsabs.harvard.edu/abs/2023A&A...671A.119C}
}

@ARTICLE{Marcy2000,
       author = {{Marcy}, Geoffrey W. and {Butler}, R. Paul},
        title = "{Planets Orbiting Other Suns}",
      journal = {\pasp},
         year = 2000,
        month = feb,
       volume = {112},
       number = {768},
        pages = {137-140},
          doi = {10.1086/316516},
       adsurl = {https://ui.adsabs.harvard.edu/abs/2000PASP..112..137M}
}

@ARTICLE{Grether2006,
       author = {{Grether}, Daniel and {Lineweaver}, Charles H.},
        title = "{How Dry is the Brown Dwarf Desert? Quantifying the Relative Number of Planets, Brown Dwarfs, and Stellar Companions around Nearby Sun-like Stars}",
      journal = {\apj},
         year = 2006,
        month = apr,
       volume = {640},
       number = {2},
        pages = {1051-1062},
          doi = {10.1086/500161},
archivePrefix = {arXiv},
       eprint = {astro-ph/0412356},
 primaryClass = {astro-ph},
       adsurl = {https://ui.adsabs.harvard.edu/abs/2006ApJ...640.1051G}
}

@ARTICLE{Vandekamp1975,
       author = {{van de Kamp}, P.},
        title = "{Unseen astrometric companions of stars.}",
      journal = {\araa},
         year = 1975,
        month = jan,
       volume = {13},
        pages = {295-333},
          doi = {10.1146/annurev.aa.13.090175.001455},
       adsurl = {https://ui.adsabs.harvard.edu/abs/1975ARA&A..13..295V}
}

@ARTICLE{Marcussen2026,
       author = {{Marcussen}, Marcus L. and {Lund}, Mikkel N. and {Grundahl}, Frank and {Huber}, Daniel and {Knudstrup}, Emil and {Kraus}, Adam L. and {Baranec}, Christoph and {Pall{\'e}}, Pere L. and {Dupuy}, Trent J. and {Huber}, Guillaume and {Ou}, James and {Werber}, Zach and {Zhang}, Ruihan and {Riddle}, Reed},
        title = "{The {\textmu} Herculis system solved after nearly three centuries}",
      journal = {\aap},
         year = 2026,
        month = may,
       volume = {709},
          eid = {A145},
        pages = {A145},
          doi = {10.1051/0004-6361/202658981},
archivePrefix = {arXiv},
       eprint = {2604.12492},
 primaryClass = {astro-ph.SR},
       adsurl = {https://ui.adsabs.harvard.edu/abs/2026A&A...709A.145M}
}

@ARTICLE{Christian2022,
       author = {{Christian}, Sam and {Vanderburg}, Andrew and {Becker}, Juliette and {Yahalomi}, Daniel A. and {Pearce}, Logan and {Zhou}, George and {Collins}, Karen A. and {Kraus}, Adam L. and {Stassun}, Keivan G. and {de Beurs}, Zoe and {Ricker}, George R. and {Vanderspek}, Roland K. and {Latham}, David W. and {Winn}, Joshua N. and {Seager}, S. and {Jenkins}, Jon M. and {Abe}, Lyu and {Agabi}, Karim and {Amado}, Pedro J. and {Baker}, David and {Barkaoui}, Khalid and {Benkhaldoun}, Zouhair and {Benni}, Paul and {Berberian}, John and {Berlind}, Perry and {Bieryla}, Allyson and {Esparza-Borges}, Emma and {Bowen}, Michael and {Brown}, Peyton and {Buchhave}, Lars A. and {Burke}, Christopher J. and {Buttu}, Marco and {Cadieux}, Charles and {Caldwell}, Douglas A. and {Charbonneau}, David and {Chazov}, Nikita and {Chimaladinne}, Sudhish and {Collins}, Kevin I. and {Combs}, Deven and {Conti}, Dennis M. and {Crouzet}, Nicolas and {de Leon}, Jerome P. and {Deljookorani}, Shila and {Diamond}, Brendan and {Doyon}, Ren{\'e} and {Dragomir}, Diana and {Dransfield}, Georgina and {Essack}, Zahra and {Evans}, Phil and {Fukui}, Akihiko and {Gan}, Tianjun and {Esquerdo}, Gilbert A. and {Gillon}, Micha{\"e}l and {Girardin}, Eric and {Guerra}, Pere and {Guillot}, Tristan and {K. Habich}, Eleanor Kate and {Henriksen}, Andreea and {Hoch}, Nora and {Isogai}, Keisuke I. and {Jehin}, Emmanu{\"e}l and {Jensen}, Eric L.~N. and {Johnson}, Marshall C. and {Livingston}, John H. and {Kielkopf}, John F. and {Kim}, Kingsley and {Kawauchi}, Kiyoe and {Krushinsky}, Vadim and {Kunzle}, Veronica and {Laloum}, Didier and {Leger}, Dominic and {Lewin}, Pablo and {Mallia}, Franco and {Massey}, Bob and {Mori}, Mayuko and {McLeod}, Kim K. and {M{\'e}karnia}, Djamel and {Mireles}, Ismael and {Mishevskiy}, Nikolay and {Tamura}, Motohide and {Murgas}, Felipe and {Narita}, Norio and {Naves}, Ramon and {Nelson}, Peter and {Osborn}, Hugh P. and {Palle}, Enric and {Parviainen}, Hannu and {Plavchan}, Peter and {Pozuelos}, Francisco J. and {Rabus}, Markus and {Relles}, Howard M. and {Rodr{\'\i}guez L{\'o}pez}, Cristina and {Quinn}, Samuel N. and {Schmider}, Francois-Xavier and {Schlieder}, Joshua E. and {Schwarz}, Richard P. and {Shporer}, Avi and {Sibbald}, Laurie and {Srdoc}, Gregor and {Stibbards}, Caitlin and {Stickler}, Hannah and {Suarez}, Olga and {Stockdale}, Chris and {Tan}, Thiam-Guan and {Terada}, Yuka and {Triaud}, Amaury and {Tronsgaard}, Rene and {Waalkes}, William C. and {Wang}, Gavin and {Watanabe}, Noriharu and {Wenceslas}, Marie-Sainte and {Wingham}, Geof and {Wittrock}, Justin and {Ziegler}, Carl},
        title = "{A Possible Alignment Between the Orbits of Planetary Systems and their Visual Binary Companions}",
      journal = {\aj},
         year = 2022,
        month = may,
       volume = {163},
       number = {5},
          eid = {207},
        pages = {207},
          doi = {10.3847/1538-3881/ac517f},
archivePrefix = {arXiv},
       eprint = {2202.00042},
 primaryClass = {astro-ph.EP},
       adsurl = {https://ui.adsabs.harvard.edu/abs/2022AJ....163..207C}
}

@ARTICLE{Behmard2022,
       author = {{Behmard}, Aida and {Dai}, Fei and {Howard}, Andrew W.},
        title = "{Stellar Companions to TESS Objects of Interest: A Test of Planet-Companion Alignment}",
      journal = {\aj},
         year = 2022,
        month = apr,
       volume = {163},
       number = {4},
          eid = {160},
        pages = {160},
          doi = {10.3847/1538-3881/ac53a7},
archivePrefix = {arXiv},
       eprint = {2202.01798},
 primaryClass = {astro-ph.EP},
       adsurl = {https://ui.adsabs.harvard.edu/abs/2022AJ....163..160B}
}

@ARTICLE{Stevenson2023,
       author = {{Stevenson}, Adam T. and {Haswell}, Carole A. and {Barnes}, John R. and {Barstow}, Joanna K.},
        title = "{Combing the brown dwarf desert with Gaia DR3}",
      journal = {\mnras},
         year = 2023,
        month = dec,
       volume = {526},
       number = {4},
        pages = {5155-5171},
          doi = {10.1093/mnras/stad3041},
archivePrefix = {arXiv},
       eprint = {2310.02695},
 primaryClass = {astro-ph.SR},
       adsurl = {https://ui.adsabs.harvard.edu/abs/2023MNRAS.526.5155S}
}

@ARTICLE{Akeson2013,
       author = {{Akeson}, R.~L. and {Chen}, X. and {Ciardi}, D. and {Crane}, M. and {Good}, J. and {Harbut}, M. and {Jackson}, E. and {Kane}, S.~R. and {Laity}, A.~C. and {Leifer}, S. and {Lynn}, M. and {McElroy}, D.~L. and {Papin}, M. and {Plavchan}, P. and {Ram{\'\i}rez}, S.~V. and {Rey}, R. and {von Braun}, K. and {Wittman}, M. and {Abajian}, M. and {Ali}, B. and {Beichman}, C. and {Beekley}, A. and {Berriman}, G.~B. and {Berukoff}, S. and {Bryden}, G. and {Chan}, B. and {Groom}, S. and {Lau}, C. and {Payne}, A.~N. and {Regelson}, M. and {Saucedo}, M. and {Schmitz}, M. and {Stauffer}, J. and {Wyatt}, P. and {Zhang}, A.},
        title = "{The NASA Exoplanet Archive: Data and Tools for Exoplanet Research}",
      journal = {\pasp},
         year = 2013,
        month = aug,
       volume = {125},
       number = {930},
        pages = {989},
          doi = {10.1086/672273},
archivePrefix = {arXiv},
       eprint = {1307.2944},
 primaryClass = {astro-ph.IM},
       adsurl = {https://ui.adsabs.harvard.edu/abs/2013PASP..125..989A}
}

@ARTICLE{2014Ma,
       author = {{Ma}, Bo and {Ge}, Jian},
        title = "{Statistical properties of brown dwarf companions: implications for different formation mechanisms}",
      journal = {\mnras},
         year = 2014,
        month = apr,
       volume = {439},
       number = {3},
        pages = {2781-2789},
          doi = {10.1093/mnras/stu134},
archivePrefix = {arXiv},
       eprint = {1303.6442},
 primaryClass = {astro-ph.EP},
       adsurl = {https://ui.adsabs.harvard.edu/abs/2014MNRAS.439.2781M}
}

@ARTICLE{Yee2017,
       author = {{Yee}, Samuel W. and {Petigura}, Erik A. and {von Braun}, Kaspar},
        title = "{Precision Stellar Characterization of FGKM Stars using an Empirical Spectral Library}",
      journal = {\apj},
         year = 2017,
        month = feb,
       volume = {836},
       number = {1},
          eid = {77},
        pages = {77},
          doi = {10.3847/1538-4357/836/1/77},
archivePrefix = {arXiv},
       eprint = {1701.00922},
 primaryClass = {astro-ph.SR},
       adsurl = {https://ui.adsabs.harvard.edu/abs/2017ApJ...836...77Y}
}

@ARTICLE{Jeffreys1946,
       author = {{Jeffreys}, Harold},
        title = "{An Invariant Form for the Prior Probability in Estimation Problems}",
      journal = {Proceedings of the Royal Society of London Series A},
         year = 1946,
        month = sep,
       volume = {186},
       number = {1007},
        pages = {453-461},
          doi = {10.1098/rspa.1946.0056},
       adsurl = {https://ui.adsabs.harvard.edu/abs/1946RSPSA.186..453J}
}

@ARTICLE{DR3Binarystar,
       author = {{Halbwachs}, Jean-Louis and {Pourbaix}, Dimitri and {Arenou}, Fr{\'e}d{\'e}ric and {Galluccio}, Laurent and {Guillout}, Patrick and {Bauchet}, Nathalie and {Marchal}, Olivier and {Sadowski}, Gilles and {Teyssier}, David},
        title = "{Gaia Data Release 3. Astrometric binary star processing}",
      journal = {\aap},
         year = 2023,
        month = jun,
       volume = {674},
          eid = {A9},
        pages = {A9},
          doi = {10.1051/0004-6361/202243969},
archivePrefix = {arXiv},
       eprint = {2206.05726},
 primaryClass = {astro-ph.SR},
       adsurl = {https://ui.adsabs.harvard.edu/abs/2023A&A...674A...9H}
}

@ARTICLE{DR3Teaser,
       author = {{Gaia Collaboration} and {Arenou}, F. and {Babusiaux}, C. and {Barstow}, M.~A. and {Faigler}, S. and {Jorissen}, A. and {Kervella}, P. and {Mazeh}, T. and {Mowlavi}, N. and {Panuzzo}, P. and {Sahlmann}, J. and {Shahaf}, S. and {Sozzetti}, A. and {Bauchet}, N. and {Damerdji}, Y. and {Gavras}, P. and {Giacobbe}, P. and {Gosset}, E. and {Halbwachs}, J.-L. and {Holl}, B. and {Lattanzi}, M.~G. and {Leclerc}, N. and {Morel}, T. and {Pourbaix}, D. and {Re Fiorentin}, P. and {Sadowski}, G. and {S{\'e}gransan}, D. and {Siopis}, C. and {Teyssier}, D. and {Zwitter}, T. and {Planquart}, L. and {Brown}, A.~G.~A. and {Vallenari}, A. and {Prusti}, T. and {de Bruijne}, J.~H.~J. and {Biermann}, M. and {Creevey}, O.~L. and {Ducourant}, C. and {Evans}, D.~W. and {Eyer}, L. and {Guerra}, R. and {Hutton}, A. and {Jordi}, C. and {Klioner}, S.~A. and {Lammers}, U.~L. and {Lindegren}, L. and {Luri}, X. and {Mignard}, F. and {Panem}, C. and {Randich}, S. and {Sartoretti}, P. and {Soubiran}, C. and {Tanga}, P. and {Walton}, N.~A. and {Bailer-Jones}, C.~A.~L. and {Bastian}, U. and {Drimmel}, R. and {Jansen}, F. and {Katz}, D. and {van Leeuwen}, F. and {Bakker}, J. and {Cacciari}, C. and {Casta{\~n}eda}, J. and {De Angeli}, F. and {Fabricius}, C. and {Fouesneau}, M. and {Fr{\'e}mat}, Y. and {Galluccio}, L. and {Guerrier}, A. and {Heiter}, U. and {Masana}, E. and {Messineo}, R. and {Nicolas}, C. and {Nienartowicz}, K. and {Pailler}, F. and {Riclet}, F. and {Roux}, W. and {Seabroke}, G.~M. and {Sordo}, R. and {Th{\'e}venin}, F. and {Gracia-Abril}, G. and {Portell}, J. and {Altmann}, M. and {Andrae}, R. and {Audard}, M. and {Bellas-Velidis}, I. and {Benson}, K. and {Berthier}, J. and {Blomme}, R. and {Burgess}, P.~W. and {Busonero}, D. and {Busso}, G. and {C{\'a}novas}, H. and {Carry}, B. and {Cellino}, A. and {Cheek}, N. and {Clementini}, G. and {Davidson}, M. and {de Teodoro}, P. and {Nu{\~n}ez Campos}, M. and {Delchambre}, L. and {Dell'Oro}, A. and {Esquej}, P. and {Fern{\'a}ndez-Hern{\'a}ndez}, J. and {Fraile}, E. and {Garabato}, D. and {Garc{\'\i}a-Lario}, P. and {Haigron}, R. and {Hambly}, N.~C. and {Harrison}, D.~L. and {Hern{\'a}ndez}, J. and {Hestroffer}, D. and {Hodgkin}, S.~T. and {Jan{\ss}en}, K. and {Jevardat de Fombelle}, G. and {Jordan}, S. and {Krone-Martins}, A. and {Lanzafame}, A.~C. and {L{\"o}ffler}, W. and {Marchal}, O. and {Marrese}, P.~M. and {Moitinho}, A. and {Muinonen}, K. and {Osborne}, P. and {Pancino}, E. and {Pauwels}, T. and {Recio-Blanco}, A. and {Reyl{\'e}}, C. and {Riello}, M. and {Rimoldini}, L. and {Roegiers}, T. and {Rybizki}, J. and {Sarro}, L.~M. and {Smith}, M. and {Utrilla}, E. and {van Leeuwen}, M. and {Abbas}, U. and {{\'A}brah{\'a}m}, P. and {Abreu Aramburu}, A. and {Aerts}, C. and {Aguado}, J.~J. and {Ajaj}, M. and {Aldea-Montero}, F. and {Altavilla}, G. and {{\'A}lvarez}, M.~A. and {Alves}, J. and {Anders}, F. and {Anderson}, R.~I. and {Anglada Varela}, E. and {Antoja}, T. and {Baines}, D. and {Baker}, S.~G. and {Balaguer-N{\'u}{\~n}ez}, L. and {Balbinot}, E. and {Balog}, Z. and {Barache}, C. and {Barbato}, D. and {Barros}, M. and {Bartolom{\'e}}, S. and {Bassilana}, J.-L. and {Becciani}, U. and {Bellazzini}, M. and {Berihuete}, A. and {Bernet}, M. and {Bertone}, S. and {Bianchi}, L. and {Binnenfeld}, A. and {Blanco-Cuaresma}, S. and {Blazere}, A. and {Boch}, T. and {Bombrun}, A. and {Bossini}, D. and {Bouquillon}, S. and {Bragaglia}, A. and {Bramante}, L. and {Breedt}, E. and {Bressan}, A. and {Brouillet}, N. and {Brugaletta}, E. and {Bucciarelli}, B. and {Burlacu}, A. and {Butkevich}, A.~G. and {Buzzi}, R. and {Caffau}, E. and {Cancelliere}, R. and {Cantat-Gaudin}, T. and {Carballo}, R. and {Carlucci}, T. and {Carnerero}, M.~I. and {Carrasco}, J.~M. and {Casamiquela}, L. and {Castellani}, M. and {Castro-Ginard}, A. and {Chaoul}, L. and {Charlot}, P. and {Chemin}, L. and {Chiaramida}, V. and {Chiavassa}, A. and {Chornay}, N. and {Comoretto}, G.},
        title = "{Gaia Data Release 3. Stellar multiplicity, a teaser for the hidden treasure}",
      journal = {\aap},
         year = 2023,
        month = jun,
       volume = {674},
          eid = {A34},
        pages = {A34},
          doi = {10.1051/0004-6361/202243782},
archivePrefix = {arXiv},
       eprint = {2206.05595},
 primaryClass = {astro-ph.SR},
       adsurl = {https://ui.adsabs.harvard.edu/abs/2023A&A...674A..34G}
}

@ARTICLE{Bessel1844,
       author = {{Bessel}, F.~W.},
        title = "{On the variations of the proper motions of Procyon and Sirius}",
      journal = {\mnras},
         year = 1844,
        month = dec,
       volume = {6},
        pages = {136-141},
          doi = {10.1093/mnras/6.11.136},
       adsurl = {https://ui.adsabs.harvard.edu/abs/1844MNRAS...6R.136B}
}

@ARTICLE{Prusti+2016,
       author = {{Gaia Collaboration} and {Prusti}, T. and {de Bruijne}, J.~H.~J. and {Brown}, A.~G.~A. and {Vallenari}, A. and {Babusiaux}, C. and {Bailer-Jones}, C.~A.~L. and {Bastian}, U. and {Biermann}, M. and {Evans}, D.~W. and {Eyer}, L. and {Jansen}, F. and {Jordi}, C. and {Klioner}, S.~A. and {Lammers}, U. and {Lindegren}, L. and {Luri}, X. and {Mignard}, F. and {Milligan}, D.~J. and {Panem}, C. and {Poinsignon}, V. and {Pourbaix}, D. and {Randich}, S. and {Sarri}, G. and {Sartoretti}, P. and {Siddiqui}, H.~I. and {Soubiran}, C. and {Valette}, V. and {van Leeuwen}, F. and {Walton}, N.~A. and {Aerts}, C. and {Arenou}, F. and {Cropper}, M. and {Drimmel}, R. and {H{\o}g}, E. and {Katz}, D. and {Lattanzi}, M.~G. and {O'Mullane}, W. and {Grebel}, E.~K. and {Holland}, A.~D. and {Huc}, C. and {Passot}, X. and {Bramante}, L. and {Cacciari}, C. and {Casta{\~n}eda}, J. and {Chaoul}, L. and {Cheek}, N. and {De Angeli}, F. and {Fabricius}, C. and {Guerra}, R. and {Hern{\'a}ndez}, J. and {Jean-Antoine-Piccolo}, A. and {Masana}, E. and {Messineo}, R. and {Mowlavi}, N. and {Nienartowicz}, K. and {Ord{\'o}{\~n}ez-Blanco}, D. and {Panuzzo}, P. and {Portell}, J. and {Richards}, P.~J. and {Riello}, M. and {Seabroke}, G.~M. and {Tanga}, P. and {Th{\'e}venin}, F. and {Torra}, J. and {Els}, S.~G. and {Gracia-Abril}, G. and {Comoretto}, G. and {Garcia-Reinaldos}, M. and {Lock}, T. and {Mercier}, E. and {Altmann}, M. and {Andrae}, R. and {Astraatmadja}, T.~L. and {Bellas-Velidis}, I. and {Benson}, K. and {Berthier}, J. and {Blomme}, R. and {Busso}, G. and {Carry}, B. and {Cellino}, A. and {Clementini}, G. and {Cowell}, S. and {Creevey}, O. and {Cuypers}, J. and {Davidson}, M. and {De Ridder}, J. and {de Torres}, A. and {Delchambre}, L. and {Dell'Oro}, A. and {Ducourant}, C. and {Fr{\'e}mat}, Y. and {Garc{\'\i}a-Torres}, M. and {Gosset}, E. and {Halbwachs}, J.-L. and {Hambly}, N.~C. and {Harrison}, D.~L. and {Hauser}, M. and {Hestroffer}, D. and {Hodgkin}, S.~T. and {Huckle}, H.~E. and {Hutton}, A. and {Jasniewicz}, G. and {Jordan}, S. and {Kontizas}, M. and {Korn}, A.~J. and {Lanzafame}, A.~C. and {Manteiga}, M. and {Moitinho}, A. and {Muinonen}, K. and {Osinde}, J. and {Pancino}, E. and {Pauwels}, T. and {Petit}, J.-M. and {Recio-Blanco}, A. and {Robin}, A.~C. and {Sarro}, L.~M. and {Siopis}, C. and {Smith}, M. and {Smith}, K.~W. and {Sozzetti}, A. and {Thuillot}, W. and {van Reeven}, W. and {Viala}, Y. and {Abbas}, U. and {Abreu Aramburu}, A. and {Accart}, S. and {Aguado}, J.~J. and {Allan}, P.~M. and {Allasia}, W. and {Altavilla}, G. and {{\'A}lvarez}, M.~A. and {Alves}, J. and {Anderson}, R.~I. and {Andrei}, A.~H. and {Anglada Varela}, E. and {Antiche}, E. and {Antoja}, T. and {Ant{\'o}n}, S. and {Arcay}, B. and {Atzei}, A. and {Ayache}, L. and {Bach}, N. and {Baker}, S.~G. and {Balaguer-N{\'u}{\~n}ez}, L. and {Barache}, C. and {Barata}, C. and {Barbier}, A. and {Barblan}, F. and {Baroni}, M. and {Barrado y Navascu{\'e}s}, D. and {Barros}, M. and {Barstow}, M.~A. and {Becciani}, U. and {Bellazzini}, M. and {Bellei}, G. and {Bello Garc{\'\i}a}, A. and {Belokurov}, V. and {Bendjoya}, P. and {Berihuete}, A. and {Bianchi}, L. and {Bienaym{\'e}}, O. and {Billebaud}, F. and {Blagorodnova}, N. and {Blanco-Cuaresma}, S. and {Boch}, T. and {Bombrun}, A. and {Borrachero}, R. and {Bouquillon}, S. and {Bourda}, G. and {Bouy}, H. and {Bragaglia}, A. and {Breddels}, M.~A. and {Brouillet}, N. and {Br{\"u}semeister}, T. and {Bucciarelli}, B. and {Budnik}, F. and {Burgess}, P. and {Burgon}, R. and {Burlacu}, A. and {Busonero}, D. and {Buzzi}, R. and {Caffau}, E. and {Cambras}, J. and {Campbell}, H. and {Cancelliere}, R. and {Cantat-Gaudin}, T. and {Carlucci}, T. and {Carrasco}, J.~M. and {Castellani}, M. and {Charlot}, P. and {Charnas}, J. and {Charvet}, P. and {Chassat}, F. and {Chiavassa}, A. and {Clotet}, M. and {Cocozza}, G. and {Collins}, R.~S. and {Collins}, P. and {Costigan}, G.},
        title = "{The Gaia mission}",
      journal = {\aap},
         year = 2016,
        month = nov,
       volume = {595},
          eid = {A1},
        pages = {A1},
          doi = {10.1051/0004-6361/201629272},
archivePrefix = {arXiv},
       eprint = {1609.04153},
 primaryClass = {astro-ph.IM},
       adsurl = {https://ui.adsabs.harvard.edu/abs/2016A&A...595A...1G}
}

@ARTICLE{El-Badry2024,
       author = {{El-Badry}, Kareem and {Lam}, Casey and {Holl}, Berry and {Halbwachs}, Jean-Louis and {Rix}, Hans-Walter and {Mazeh}, Tsevi and {Shahaf}, Sahar},
        title = "{A generative model for Gaia astrometric orbit catalogs: selection functions for binary stars, giant planets, and compact object companions}",
      journal = {The Open Journal of Astrophysics},
         year = 2024,
        month = nov,
       volume = {7},
          eid = {100},
        pages = {100},
          doi = {10.33232/001c.125461},
archivePrefix = {arXiv},
       eprint = {2411.00088},
 primaryClass = {astro-ph.SR},
       adsurl = {https://ui.adsabs.harvard.edu/abs/2024OJAp....7E.100E}
}

@ARTICLE{Zechmeister2018,
       author = {{Zechmeister}, M. and {Reiners}, A. and {Amado}, P.~J. and {Azzaro}, M. and {Bauer}, F.~F. and {B{\'e}jar}, V.~J.~S. and {Caballero}, J.~A. and {Guenther}, E.~W. and {Hagen}, H.-J. and {Jeffers}, S.~V. and {Kaminski}, A. and {K{\"u}rster}, M. and {Launhardt}, R. and {Montes}, D. and {Morales}, J.~C. and {Quirrenbach}, A. and {Reffert}, S. and {Ribas}, I. and {Seifert}, W. and {Tal-Or}, L. and {Wolthoff}, V.},
        title = "{Spectrum radial velocity analyser (SERVAL). High-precision radial velocities and two alternative spectral indicators}",
      journal = {\aap},
         year = 2018,
        month = jan,
       volume = {609},
          eid = {A12},
        pages = {A12},
          doi = {10.1051/0004-6361/201731483},
archivePrefix = {arXiv},
       eprint = {1710.10114},
 primaryClass = {astro-ph.IM},
       adsurl = {https://ui.adsabs.harvard.edu/abs/2018A&A...609A..12Z}
}

@INPROCEEDINGS{Kurucz1992,
       author = {{Kurucz}, R.~L.},
        title = "{Model Atmospheres for Population Synthesis}",
    booktitle = {The Stellar Populations of Galaxies},
         year = 1992,
       editor = {{Barbuy}, Beatriz and {Renzini}, Alvio},
       series = {IAU Symposium},
       volume = {149},
        month = jan,
        pages = {225},
       adsurl = {https://ui.adsabs.harvard.edu/abs/1992IAUS..149..225K}
}

@ARTICLE{Perryman2014,
       author = {{Perryman}, Michael and {Hartman}, Joel and {Bakos}, G{\'a}sp{\'a}r {\'A}. and {Lindegren}, Lennart},
        title = "{Astrometric Exoplanet Detection with Gaia}",
      journal = {\apj},
         year = 2014,
        month = dec,
       volume = {797},
       number = {1},
          eid = {14},
        pages = {14},
          doi = {10.1088/0004-637X/797/1/14},
archivePrefix = {arXiv},
       eprint = {1411.1173},
 primaryClass = {astro-ph.EP},
       adsurl = {https://ui.adsabs.harvard.edu/abs/2014ApJ...797...14P}
}

@ARTICLE{Lammers2026,
       author = {{Lammers}, Caleb and {Winn}, Joshua N.},
        title = "{On the Exoplanet Yield of Gaia Astrometry}",
      journal = {\aj},
         year = 2026,
        month = jan,
       volume = {171},
       number = {1},
          eid = {18},
        pages = {18},
          doi = {10.3847/1538-3881/ae21de},
archivePrefix = {arXiv},
       eprint = {2511.04673},
 primaryClass = {astro-ph.EP},
       adsurl = {https://ui.adsabs.harvard.edu/abs/2026AJ....171...18L}
}

@ARTICLE{Sozzetti2023,
       author = {{Sozzetti}, A. and {Pinamonti}, M. and {Damasso}, M. and {Desidera}, S. and {Biazzo}, K. and {Bonomo}, A.~S. and {Nardiello}, D. and {Gratton}, R. and {Lanza}, A.~F. and {Malavolta}, L. and {Giacobbe}, P. and {Affer}, L. and {Bignamini}, A. and {Borsa}, F. and {Boschin}, W. and {Brogi}, M. and {Cabona}, L. and {Claudi}, R. and {Covino}, E. and {Di Fabrizio}, L. and {Ghedina}, A. and {Harutyunyan}, A. and {Knapic}, C. and {Maldonado}, J. and {Maggio}, A. and {Mancini}, L. and {Mantovan}, G. and {Marzari}, F. and {Messina}, S. and {Micela}, G. and {Molinari}, E. and {Montalto}, M. and {Naponiello}, L. and {Pagano}, I. and {Pedani}, M. and {Piotto}, G. and {Poretti}, E. and {Scandariato}, G. and {Silvotti}, R. and {Turrini}, D.},
        title = "{The GAPS Programme at TNG. XLVII. A conundrum resolved: HIP 66074b/Gaia-3b characterised as a massive giant planet on a quasi-face-on and extremely elongated orbit}",
      journal = {\aap},
         year = 2023,
        month = sep,
       volume = {677},
          eid = {L15},
        pages = {L15},
          doi = {10.1051/0004-6361/202347329},
archivePrefix = {arXiv},
       eprint = {2307.08653},
 primaryClass = {astro-ph.EP},
       adsurl = {https://ui.adsabs.harvard.edu/abs/2023A&A...677L..15S}
}

@ARTICLE{Pinamonti2026,
       author = {{Pinamonti}, M. and {Sozzetti}, A. and {Barbato}, D. and {Desidera}, S. and {Biazzo}, K. and {Bonomo}, A.~S. and {Lanza}, A.~F. and {Naponiello}, L. and {Affer}, L. and {Anche}, R.~M. and {Andreuzzi}, G. and {Basilicata}, M. and {Brinjikji}, M. and {Brogi}, M. and {Cabona}, L. and {Carolo}, E. and {Colombo}, S. and {Damasso}, M. and {D'Arpa}, M. and {Di Filippo}, S. and {Harutyunyan}, A. and {Hom}, J. and {Mancini}, L. and {Mantovan}, G. and {Nardiello}, D. and {Santhakumari}, K.~K.~R. and {Zingales}, T.},
        title = "{The GAPS programme at TNG: LXX. HD128717 B/Gaia-6 B: A long-period eccentric low-mass brown dwarf from astrometry and radial velocities}",
      journal = {\aap},
         year = 2026,
        month = mar,
       volume = {707},
          eid = {A67},
        pages = {A67},
          doi = {10.1051/0004-6361/202557414},
archivePrefix = {arXiv},
       eprint = {2512.04606},
 primaryClass = {astro-ph.EP},
       adsurl = {https://ui.adsabs.harvard.edu/abs/2026A&A...707A..67P}
}

@ARTICLE{Fitzmaurice2024,
       author = {{Fitzmaurice}, Evan and {Stef{\'a}nsson}, Gu{\dj}mundur and {Kavanagh}, Robert D. and {Mahadevan}, Suvrath and {Ca{\~n}as}, Caleb I. and {Winn}, Joshua N. and {Robertson}, Paul and {Ninan}, Joe P. and {Albrecht}, Simon and {Callingham}, J.~R. and {Cochran}, William D. and {Delamer}, Megan and {Ford}, Eric B. and {Kanodia}, Shubham and {Lin}, Andrea S.~J. and {Marcussen}, Marcus L. and {Pope}, Benjamin J.~S. and {Ramsey}, Lawrence W. and {Roy}, Arpita and {Vedantham}, Harish and {Wright}, Jason T.},
        title = "{Astrometry and Precise Radial Velocities Yield a Complete Orbital Solution for the Nearby Eccentric Brown Dwarf LHS 1610 b}",
      journal = {\aj},
         year = 2024,
        month = sep,
       volume = {168},
       number = {3},
          eid = {140},
        pages = {140},
          doi = {10.3847/1538-3881/ad57be},
archivePrefix = {arXiv},
       eprint = {2310.07827},
 primaryClass = {astro-ph.SR},
       adsurl = {https://ui.adsabs.harvard.edu/abs/2024AJ....168..140F}
}

@ARTICLE{Winn2022,
       author = {{Winn}, Joshua N.},
        title = "{Joint Constraints on Exoplanetary Orbits from Gaia DR3 and Doppler Data}",
      journal = {\aj},
         year = 2022,
        month = nov,
       volume = {164},
       number = {5},
          eid = {196},
        pages = {196},
          doi = {10.3847/1538-3881/ac9126},
archivePrefix = {arXiv},
       eprint = {2209.05516},
 primaryClass = {astro-ph.EP},
       adsurl = {https://ui.adsabs.harvard.edu/abs/2022AJ....164..196W}
}

@ARTICLE{MarcussenAlbrecht2023,
       author = {{Marcussen}, Marcus L. and {Albrecht}, Simon H.},
        title = "{Spectroscopic Follow-up of Gaia Exoplanet Candidates: Impostor Binary Stars Invade the Gaia DR3 Astrometric Exoplanet Candidates}",
      journal = {\aj},
         year = 2023,
        month = jun,
       volume = {165},
       number = {6},
          eid = {266},
        pages = {266},
          doi = {10.3847/1538-3881/acd53d},
archivePrefix = {arXiv},
       eprint = {2305.08623},
 primaryClass = {astro-ph.EP},
       adsurl = {https://ui.adsabs.harvard.edu/abs/2023AJ....165..266M}
}

@ARTICLE{Stefansson2025,
       author = {{Stef{\'a}nsson}, Gudmundur and {Mahadevan}, Suvrath and {Winn}, Joshua N. and {Marcussen}, Marcus L. and {Kanodia}, Shubham and {Albrecht}, Simon and {Fitzmaurice}, Evan and {Mikulskyt{\.{e}}}, On{\.{e}} and {Ca{\~n}as}, Caleb I. and {Espinoza-Retamal}, Juan I. and {Zwart}, Yiri and {Krolikowski}, Daniel M. and {Hotnisky}, Andrew and {Robertson}, Paul and {Alvarado-Montes}, Jaime A. and {Bender}, Chad F. and {Blake}, Cullen H. and {Callingham}, J.~R. and {Cochran}, William D. and {Delamer}, Megan and {Diddams}, Scott A. and {Dong}, Jiayin and {Fernandes}, Rachel B. and {Giovinazzi}, Mark R. and {Halverson}, Samuel and {Libby-Roberts}, Jessica and {Logsdon}, Sarah E. and {McElwain}, Michael W. and {Ninan}, Joe P. and {Rajagopal}, Jayadev and {Reji}, Varghese and {Roy}, Arpita and {Schwab}, Christian and {Wright}, Jason T.},
        title = "{Gaia-4b and 5b: Radial Velocity Confirmation of Gaia Astrometric Orbital Solutions Reveal a Massive Planet and a Brown Dwarf Orbiting Low-mass Stars}",
      journal = {\aj},
         year = 2025,
        month = feb,
       volume = {169},
       number = {2},
          eid = {107},
        pages = {107},
          doi = {10.3847/1538-3881/ada9e1},
archivePrefix = {arXiv},
       eprint = {2410.05654},
 primaryClass = {astro-ph.EP},
       adsurl = {https://ui.adsabs.harvard.edu/abs/2025AJ....169..107S}
}

@INPROCEEDINGS{NOT2010,
       author = {{Djupvik}, Anlaug Amanda and {Andersen}, Johannes},
        title = "{The Nordic Optical Telescope}",
    booktitle = {Highlights of Spanish Astrophysics V},
         year = 2010,
       series = {Astrophysics and Space Science Proceedings},
       volume = {14},
        month = jan,
        pages = {211},
          doi = {10.1007/978-3-642-11250-8_21},
archivePrefix = {arXiv},
       eprint = {0901.4015},
 primaryClass = {astro-ph.IM},
       adsurl = {https://ui.adsabs.harvard.edu/abs/2010ASSP...14..211D}
}

@INPROCEEDINGS{Rucinski1999,
       author = {{Rucinski}, S.},
        title = "{Determination of Broadening Functions Using the Singular-Value Decomposition (SVD) Technique}",
    booktitle = {IAU Colloquium 170: Precise Stellar Radial Velocities},
         year = 1999,
       editor = {{Hearnshaw}, J.~B. and {Scarfe}, Colin David},
       series = {Astronomical Society of the Pacific Conference Series},
       volume = {185},
        month = jan,
        pages = {82},
          doi = {10.48550/arXiv.astro-ph/9807327},
archivePrefix = {arXiv},
       eprint = {astro-ph/9807327},
 primaryClass = {astro-ph},
       adsurl = {https://ui.adsabs.harvard.edu/abs/1999ASPC..185...82R}
}

@ARTICLE{Foreman-Mackey2013,
       author = {{Foreman-Mackey}, Daniel and {Hogg}, David W. and {Lang}, Dustin and {Goodman}, Jonathan},
        title = "{emcee: The MCMC Hammer}",
      journal = {\pasp},
         year = 2013,
        month = mar,
       volume = {125},
       number = {925},
        pages = {306},
          doi = {10.1086/670067},
archivePrefix = {arXiv},
       eprint = {1202.3665},
 primaryClass = {astro-ph.IM},
       adsurl = {https://ui.adsabs.harvard.edu/abs/2013PASP..125..306F}
}

@ARTICLE{Householder2022,
       author = {{Householder}, Aaron and {Weiss}, Lauren},
        title = "{The Inconsistent use of $\omega$ in the RV Equation}",
      journal = {arXiv e-prints},
         year = 2022,
        month = dec,
          eid = {arXiv:2212.06966},
        pages = {arXiv:2212.06966},
          doi = {10.48550/arXiv.2212.06966},
archivePrefix = {arXiv},
       eprint = {2212.06966},
 primaryClass = {astro-ph.EP},
       adsurl = {https://ui.adsabs.harvard.edu/abs/2022arXiv221206966H}
}

@ARTICLE{Lindegren2021,
       author = {{Lindegren}, L. and {Klioner}, S.~A. and {Hern{\'a}ndez}, J. and {Bombrun}, A. and {Ramos-Lerate}, M. and {Steidelm{\"u}ller}, H. and {Bastian}, U. and {Biermann}, M. and {de Torres}, A. and {Gerlach}, E. and {Geyer}, R. and {Hilger}, T. and {Hobbs}, D. and {Lammers}, U. and {McMillan}, P.~J. and {Stephenson}, C.~A. and {Casta{\~n}eda}, J. and {Davidson}, M. and {Fabricius}, C. and {Gracia-Abril}, G. and {Portell}, J. and {Rowell}, N. and {Teyssier}, D. and {Torra}, F. and {Bartolom{\'e}}, S. and {Clotet}, M. and {Garralda}, N. and {Gonz{\'a}lez-Vidal}, J.~J. and {Torra}, J. and {Abbas}, U. and {Altmann}, M. and {Anglada Varela}, E. and {Balaguer-N{\'u}{\~n}ez}, L. and {Balog}, Z. and {Barache}, C. and {Becciani}, U. and {Bernet}, M. and {Bertone}, S. and {Bianchi}, L. and {Bouquillon}, S. and {Brown}, A.~G.~A. and {Bucciarelli}, B. and {Busonero}, D. and {Butkevich}, A.~G. and {Buzzi}, R. and {Cancelliere}, R. and {Carlucci}, T. and {Charlot}, P. and {Cioni}, M. -R.~L. and {Crosta}, M. and {Crowley}, C. and {del Peloso}, E.~F. and {del Pozo}, E. and {Drimmel}, R. and {Esquej}, P. and {Fienga}, A. and {Fraile}, E. and {Gai}, M. and {Garcia-Reinaldos}, M. and {Guerra}, R. and {Hambly}, N.~C. and {Hauser}, M. and {Jan{\ss}en}, K. and {Jordan}, S. and {Kostrzewa-Rutkowska}, Z. and {Lattanzi}, M.~G. and {Liao}, S. and {Licata}, E. and {Lister}, T.~A. and {L{\"o}ffler}, W. and {Marchant}, J.~M. and {Masip}, A. and {Mignard}, F. and {Mints}, A. and {Molina}, D. and {Mora}, A. and {Morbidelli}, R. and {Murphy}, C.~P. and {Pagani}, C. and {Panuzzo}, P. and {Pe{\~n}alosa Esteller}, X. and {Poggio}, E. and {Re Fiorentin}, P. and {Riva}, A. and {Sagrist{\`a} Sell{\'e}s}, A. and {Sanchez Gimenez}, V. and {Sarasso}, M. and {Sciacca}, E. and {Siddiqui}, H.~I. and {Smart}, R.~L. and {Souami}, D. and {Spagna}, A. and {Steele}, I.~A. and {Taris}, F. and {Utrilla}, E. and {van Reeven}, W. and {Vecchiato}, A.},
        title = "{Gaia Early Data Release 3. The astrometric solution}",
      journal = {\aap},
         year = 2021,
        month = may,
       volume = {649},
          eid = {A2},
        pages = {A2},
          doi = {10.1051/0004-6361/202039709},
archivePrefix = {arXiv},
       eprint = {2012.03380},
 primaryClass = {astro-ph.IM},
       adsurl = {https://ui.adsabs.harvard.edu/abs/2021A&A...649A...2L}
}

@ARTICLE{Wittenmyer2009,
       author = {{Wittenmyer}, Robert A. and {Endl}, Michael and {Cochran}, William D. and {Levison}, Harold F. and {Henry}, Gregory W.},
        title = "{A Search for Multi-Planet Systems Using the Hobby-Eberly Telescope}",
      journal = {\apjs},
         year = 2009,
        month = may,
       volume = {182},
       number = {1},
        pages = {97-119},
          doi = {10.1088/0067-0049/182/1/97},
archivePrefix = {arXiv},
       eprint = {0903.0652},
 primaryClass = {astro-ph.EP},
       adsurl = {https://ui.adsabs.harvard.edu/abs/2009ApJS..182...97W}
}

@ARTICLE{Holl2023,
       author = {{Holl}, B. and {Sozzetti}, A. and {Sahlmann}, J. and {Giacobbe}, P. and {S{\'e}gransan}, D. and {Unger}, N. and {Delisle}, J. -B. and {Barbato}, D. and {Lattanzi}, M.~G. and {Morbidelli}, R. and {Sosnowska}, D.},
        title = "{Gaia Data Release 3. Astrometric orbit determination with Markov chain Monte Carlo and genetic algorithms: Systems with stellar, sub-stellar, and planetary mass companions}",
      journal = {\aap},
         year = 2023,
        month = jun,
       volume = {674},
          eid = {A10},
        pages = {A10},
          doi = {10.1051/0004-6361/202244161},
archivePrefix = {arXiv},
       eprint = {2206.05439},
 primaryClass = {astro-ph.EP},
       adsurl = {https://ui.adsabs.harvard.edu/abs/2023A&A...674A..10H}
}

@ARTICLE{ElBadry2021,
       author = {{El-Badry}, Kareem and {Rix}, Hans-Walter and {Heintz}, Tyler M.},
        title = "{A million binaries from Gaia eDR3: sample selection and validation of Gaia parallax uncertainties}",
      journal = {\mnras},
         year = 2021,
        month = sep,
       volume = {506},
       number = {2},
        pages = {2269-2295},
          doi = {10.1093/mnras/stab323},
archivePrefix = {arXiv},
       eprint = {2101.05282},
 primaryClass = {astro-ph.SR},
       adsurl = {https://ui.adsabs.harvard.edu/abs/2021MNRAS.506.2269E}
}

@ARTICLE{2017Msngr.169...21B,
       author = {{Bouchy}, F. and {Doyon}, R. and {Artigau}, {\'E}. and {Melo}, C. and {Hernandez}, O. and {Wildi}, F. and {Delfosse}, X. and {Lovis}, C. and {Figueira}, P. and {Canto Martins}, B.~L.. and {Gonz{\'a}lez Hern{\'a}ndez}, J.~I.. and {Thibault}, S. and {Reshetov}, V. and {Pepe}, F. and {Santos}, N.~C. and {de Medeiros}, J.~R.. and {Rebolo}, R. and {Abreu}, M. and {Adibekyan}, V.~Z. and {Bandy}, T. and {Benz}, W. and {Blind}, N. and {Bohlender}, D. and {Boisse}, I. and {Bovay}, S. and {Broeg}, C. and {Brousseau}, D. and {Cabral}, A. and {Chazelas}, B. and {Cloutier}, R. and {Coelho}, J. and {Conod}, U. and {Cumming}, A. and {Delabre}, B. and {Genolet}, L. and {Hagelberg}, J. and {Jayawardhana}, R. and {K{\"a}ufl}, H.-U. and {Lafreni{\`e}re}, D. and {de Castro Le{\~a}o}, I.. and {Malo}, L. and {de Medeiros Martins}, A.. and {Matthews}, J.~M. and {Metchev}, S. and {Oshagh}, M. and {Ouellet}, M. and {Parro}, V.~C. and {Rasilla Pi{\~n}eiro}, J.~L.. and {Santos}, P. and {Sarajlic}, M. and {Segovia}, A. and {Sordet}, M. and {Udry}, S. and {Valencia}, D. and {Vall{\'e}e}, P. and {Venn}, K. and {Wade}, G.~A. and {Saddlemyer}, L.},
        title = "{Near-InfraRed Planet Searcher to Join HARPS on the ESO 3.6-metre Telescope}",
      journal = {The Messenger},
         year = 2017,
        month = sep,
       volume = {169},
        pages = {21-27},
          doi = {10.18727/0722-6691/5034},
       adsurl = {https://ui.adsabs.harvard.edu/abs/2017Msngr.169...21B}
}

@ARTICLE{2025A&A...700A..10B,
       author = {{Bouchy}, Fran{\c{c}}ois and {Doyon}, Ren{\'e} and {Pepe}, Francesco and {Melo}, Claudio and {Artigau}, {\'E}tienne and {Malo}, Lison and {Wildi}, Fran{\c{c}}ois and {Baron}, Fr{\'e}d{\'e}rique and {Delfosse}, Xavier and {De Medeiros}, Jose Renan and {Rebolo}, Rafael and {Santos}, Nuno C. and {Wade}, Gregg and {Allart}, Romain and {Al Moulla}, Khaled and {Blind}, Nicolas and {Cadieux}, Charles and {Canto Martins}, Bruno L. and {Cook}, Neil J. and {Dumusque}, Xavier and {Frensch}, Yolanda and {Genest}, Fr{\'e}d{\'e}ric and {Gonz{\'a}lez Hern{\'a}ndez}, Jonay I. and {Grieves}, Nolan and {Lo Curto}, Gaspare and {Lovis}, Christophe and {Mignon}, Lucile and {Nielsen}, Louise D. and {Poulin-Girard}, Anne-Sophie and {Rasilla}, Jos{\'e} Luis and {Reshetov}, Vladimir and {Sosnowska}, Danuta and {Sordet}, Michael and {Saint-Antoine}, Jonathan and {Su{\'a}rez Mascare{\~n}o}, Alejandro and {Thibault}, Simon and {Vall{\'e}e}, Philippe and {Vandal}, Thomas and {Abreu}, Manuel and {Aguiar}, Jos{\'e} L.~A. and {Allain}, Guillaume and {Arial}, Tomy and {Auger}, Hugues and {Barros}, Susana C.~C. and {Bazinet}, Luc and {Benneke}, Bj{\"o}rn and {Bonfils}, Xavier and {Boucher}, Anne and {Bourrier}, Vincent and {Bovay}, S{\'e}bastien and {Broeg}, Christopher and {Brousseau}, Denis and {Bruniquel}, Vincent and {Bryan}, Marta and {Cabral}, Alexandre and {Carmona}, Andres and {Carteret}, Yann and {Challita}, Zalpha and {Chazelas}, Bruno and {Cloutier}, Ryan and {Coelho}, Jo{\~a}o and {Cointepas}, Marion and {Conod}, Uriel and {Cowan}, Nicolas B. and {Cristo}, Eduardo and {Gomes da Silva}, Jo{\~a}o and {Dauplaise}, Laurie and {Darveau-Bernier}, Antoine and {de Lima Gomes}, Roseane and {de Freitas}, Daniel Brito and {Delgado-Mena}, Elisa and {Delisle}, Jean-Baptiste and {Ehrenreich}, David and {Faria}, Jo{\~a}o and {Figueira}, Pedro and {Fontinele}, Dasaev O. and {Forveille}, Thierry and {Gagn{\'e}}, Jonathan and {Genolet}, Ludovic and {T{\'e}mich}, F{\'e}lix Gracia and {Hernandez}, Olivier and {Hobson}, Melissa J. and {Hoeijmakers}, Jens and {Hubin}, Norbert and {Jahandar}, Farbod and {Jayawardhana}, Ray and {K{\"a}ufl}, Hans-Ulrich and {Kerley}, Dan and {Kolb}, Johann and {Krishnamurthy}, Vigneshwaran and {Lafreni{\`e}re}, David and {Lamontagne}, Pierrot and {Larue}, Pierre and {Leath}, Henry and {L'Heureux}, Alexandrine and {de Castro Le{\~a}o}, Izan and {Lim}, Olivia and {Martins}, Allan M. and {Matthews}, Jaymie and {Mayer}, Jean-S{\'e}bastien and {Messias}, Yuri S. and {Metchev}, Stan and {Moranta}, Leslie and {Mordasini}, Christoph and {Mounzer}, Dany and {Nari}, Nicola and {Osborn}, Ares and {Ouellet}, Mathieu and {Otegi}, Jon and {Parc}, L{\'e}na and {Pasquini}, Luca and {Passegger}, Vera M. and {Pelletier}, Stefan and {Peroux}, C{\'e}line and {Piaulet-Ghorayeb}, Caroline and {Plotnykov}, Mykhaylo and {Pompei}, Emanuela and {Rowe}, Jason and {Sarajlic}, Mirsad and {Segovia}, Alex and {Seidel}, Julia and {S{\'e}gransan}, Damien and {Schnell}, Robin and {Costa Silva}, Ana Rita and {Srivastava}, Avidaan and {Stefanov}, Atanas K. and {Teixeira}, M{\'a}rcio A. and {Udry}, St{\'e}phane and {Valencia}, Diana and {Vaulato}, Valentina and {Wardenier}, Joost P. and {Wehbe}, Bachar and {Weisserman}, Drew and {Wevers}, Ivan and {Yariv}, Vincent and {Zins}, G{\'e}rard},
        title = "{NIRPS joining HARPS at ESO 3.6 m: On-sky performance and science objectives}",
      journal = {\aap},
         year = 2025,
        month = aug,
       volume = {700},
          eid = {A10},
        pages = {A10},
          doi = {10.1051/0004-6361/202453341},
archivePrefix = {arXiv},
       eprint = {2507.21767},
 primaryClass = {astro-ph.IM},
       adsurl = {https://ui.adsabs.harvard.edu/abs/2025A&A...700A..10B}
}

@INPROCEEDINGS{2017SPIE10400E..18W,
       author = {{Wildi}, F. and {Blind}, N. and {Reshetov}, V. and {Hernandez}, O. and {Genolet}, L. and {Conod}, U. and {Sordet}, M. and {Segovilla}, A. and {Rasilla}, J.~L. and {Brousseau}, D. and {Thibault}, S. and {Delabre}, B. and {Bandy}, T. and {Sarajlic}, M. and {Cabral}, A. and {Bovay}, S. and {Vall{\'e}e}, Ph. and {Bouchy}, F. and {Doyon}, R. and {Artigau}, E. and {Pepe}, F. and {Hagelberg}, J. and {Melo}, C. and {Delfosse}, X. and {Figueira}, P. and {Santos}, N.~C. and {Gonz{\'a}lez Hern{\'a}ndez}, J.~I. and {de Medeiros}, J.~R. and {Rebolo}, R. and {Broeg}, Ch. and {Benz}, W. and {Boisse}, I. and {Malo}, L. and {K{\"a}ufl}, U. and {Saddlemyer}, L.},
        title = "{NIRPS: an adaptive-optics assisted radial velocity spectrograph to chase exoplanets around M-stars}",
    booktitle = {Society of Photo-Optical Instrumentation Engineers (SPIE) Conference Series},
         year = 2017,
       editor = {{Shaklan}, Stuart},
       series = {Society of Photo-Optical Instrumentation Engineers (SPIE) Conference Series},
       volume = {10400},
        month = sep,
          eid = {1040018},
        pages = {1040018},
          doi = {10.1117/12.2275660},
       adsurl = {https://ui.adsabs.harvard.edu/abs/2017SPIE10400E..18W}
}

@ARTICLE{Hidalgo18,
       author = {{Hidalgo}, Sebastian L. and {Pietrinferni}, Adriano and {Cassisi}, Santi and {Salaris}, Maurizio and {Mucciarelli}, Alessio and {Savino}, Alessandro and {Aparicio}, Antonio and {Silva Aguirre}, Victor and {Verma}, Kuldeep},
        title = "{The Updated BaSTI Stellar Evolution Models and Isochrones. I. Solar-scaled Calculations}",
      journal = {\apj},
         year = 2018,
        month = apr,
       volume = {856},
       number = {2},
          eid = {125},
        pages = {125},
          doi = {10.3847/1538-4357/aab158},
archivePrefix = {arXiv},
       eprint = {1802.07319},
 primaryClass = {astro-ph.GA},
       adsurl = {https://ui.adsabs.harvard.edu/abs/2018ApJ...856..125H}
}

@ARTICLE{Aguirre22,
       author = {{Aguirre B{\o}rsen-Koch}, V. and {R{\o}rsted}, J.~L. and {Justesen}, A.~B. and {Stokholm}, A. and {Verma}, K. and {Winther}, M.~L. and {Knudstrup}, E. and {Nielsen}, K.~B. and {Sahlholdt}, C. and {Larsen}, J.~R. and {Cassisi}, S. and {Serenelli}, A.~M. and {Casagrande}, L. and {Christensen-Dalsgaard}, J. and {Davies}, G.~R. and {Ferguson}, J.~W. and {Lund}, M.~N. and {Weiss}, A. and {White}, T.~R.},
        title = "{The BAyesian STellar algorithm (BASTA): a fitting tool for stellar studies, asteroseismology, exoplanets, and Galactic archaeology}",
      journal = {\mnras},
         year = 2022,
        month = jan,
       volume = {509},
       number = {3},
        pages = {4344-4364},
          doi = {10.1093/mnras/stab2911},
archivePrefix = {arXiv},
       eprint = {2109.14622},
 primaryClass = {astro-ph.SR},
       adsurl = {https://ui.adsabs.harvard.edu/abs/2022MNRAS.509.4344A}
}

@ARTICLE{Riello21,
       author = {{Riello}, M. and {De Angeli}, F. and {Evans}, D.~W. and {Montegriffo}, P. and {Carrasco}, J.~M. and {Busso}, G. and {Palaversa}, L. and {Burgess}, P.~W. and {Diener}, C. and {Davidson}, M. and {Rowell}, N. and {Fabricius}, C. and {Jordi}, C. and {Bellazzini}, M. and {Pancino}, E. and {Harrison}, D.~L. and {Cacciari}, C. and {van Leeuwen}, F. and {Hambly}, N.~C. and {Hodgkin}, S.~T. and {Osborne}, P.~J. and {Altavilla}, G. and {Barstow}, M.~A. and {Brown}, A.~G.~A. and {Castellani}, M. and {Cowell}, S. and {De Luise}, F. and {Gilmore}, G. and {Giuffrida}, G. and {Hidalgo}, S. and {Holland}, G. and {Marinoni}, S. and {Pagani}, C. and {Piersimoni}, A.~M. and {Pulone}, L. and {Ragaini}, S. and {Rainer}, M. and {Richards}, P.~J. and {Sanna}, N. and {Walton}, N.~A. and {Weiler}, M. and {Yoldas}, A.},
        title = "{Gaia Early Data Release 3. Photometric content and validation}",
      journal = {\aap},
         year = 2021,
        month = may,
       volume = {649},
          eid = {A3},
        pages = {A3},
          doi = {10.1051/0004-6361/202039587},
archivePrefix = {arXiv},
       eprint = {2012.01916},
 primaryClass = {astro-ph.IM},
       adsurl = {https://ui.adsabs.harvard.edu/abs/2021A&A...649A...3R}
}

@ARTICLE{stefansson2016,
       author = {{Stefansson}, Gudmundur and {Hearty}, Frederick and {Robertson}, Paul and {Mahadevan}, Suvrath and {Anderson}, Tyler and {Levi}, Eric and {Bender}, Chad and {Nelson}, Matthew and {Monson}, Andrew and {Blank}, Basil and {Halverson}, Samuel and {Henderson}, Chuck and {Ramsey}, Lawrence and {Roy}, Arpita and {Schwab}, Christian and {Terrien}, Ryan},
        title = "{A Versatile Technique to Enable Sub-milli-Kelvin Instrument Stability for Precise Radial Velocity Measurements: Tests with the Habitable-zone Planet Finder}",
      journal = {\apj},
         year = 2016,
        month = dec,
       volume = {833},
       number = {2},
          eid = {175},
        pages = {175},
          doi = {10.3847/1538-4357/833/2/175},
archivePrefix = {arXiv},
       eprint = {1610.06216},
 primaryClass = {astro-ph.IM},
       adsurl = {https://ui.adsabs.harvard.edu/abs/2016ApJ...833..175S}
}

\begin{appendix}
\onecolumn
\section{Additional plots}
\label{appendixA}

\begin{figure*}[ht!]
    \centering
    \includegraphics[width=18cm]{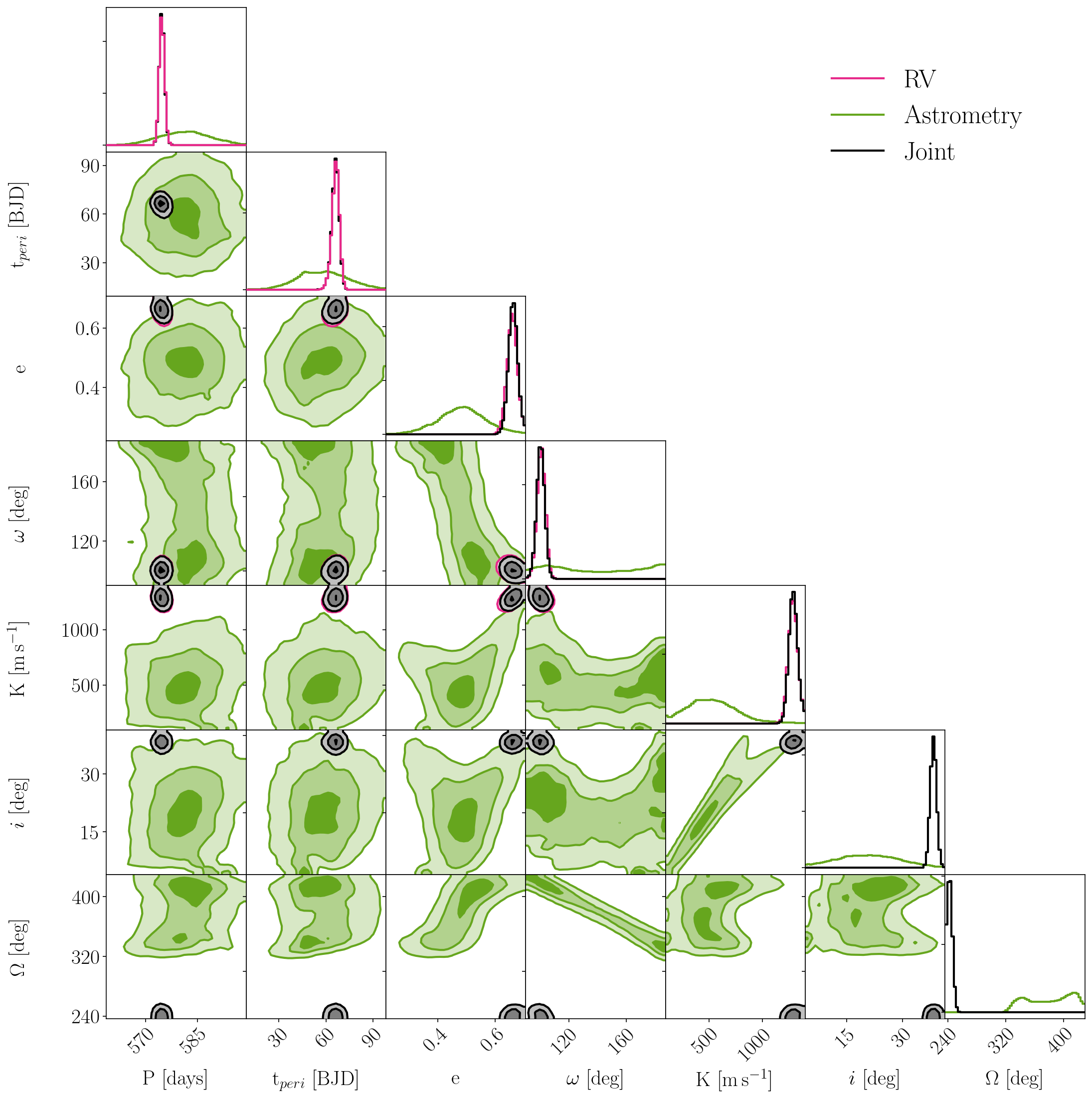}
    \caption{Corner plot of the joint posterior probability distribution for HD 5433. Shaded areas are kernel density estimates (KDEs) of the RV-only (pink), astrometry-only (green), and joint (black) posterior probability distributions.}
    \label{fig:corner7984}
\end{figure*}

\begin{figure*}
    \centering
    \includegraphics[width=18cm]{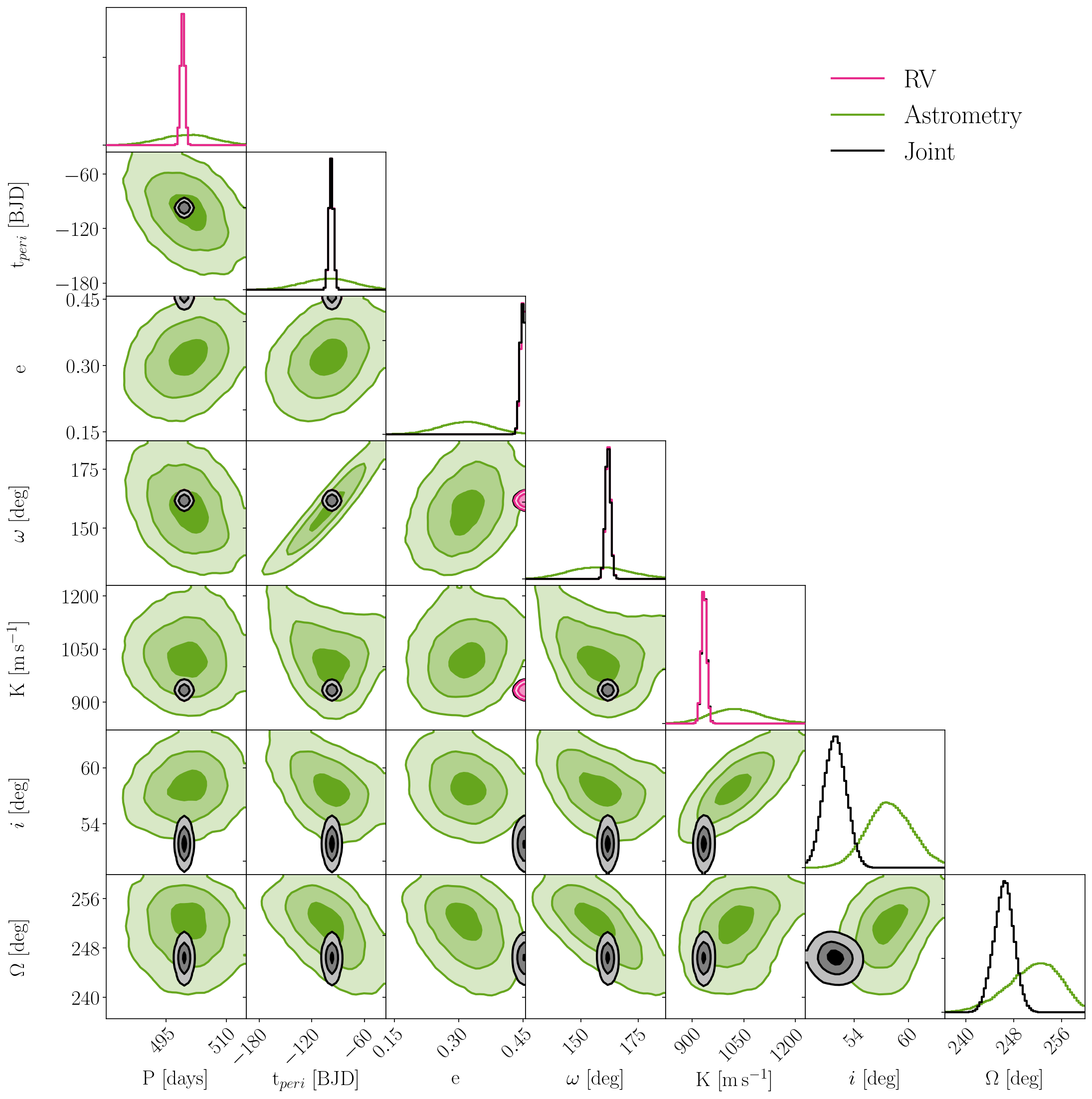}
    \caption{Same as Fig.~\ref{fig:corner7984}, but for HD 91669.}
    \label{fig:corner6128}
\end{figure*}

\begin{figure*}
    \centering
    \includegraphics[width=18cm]{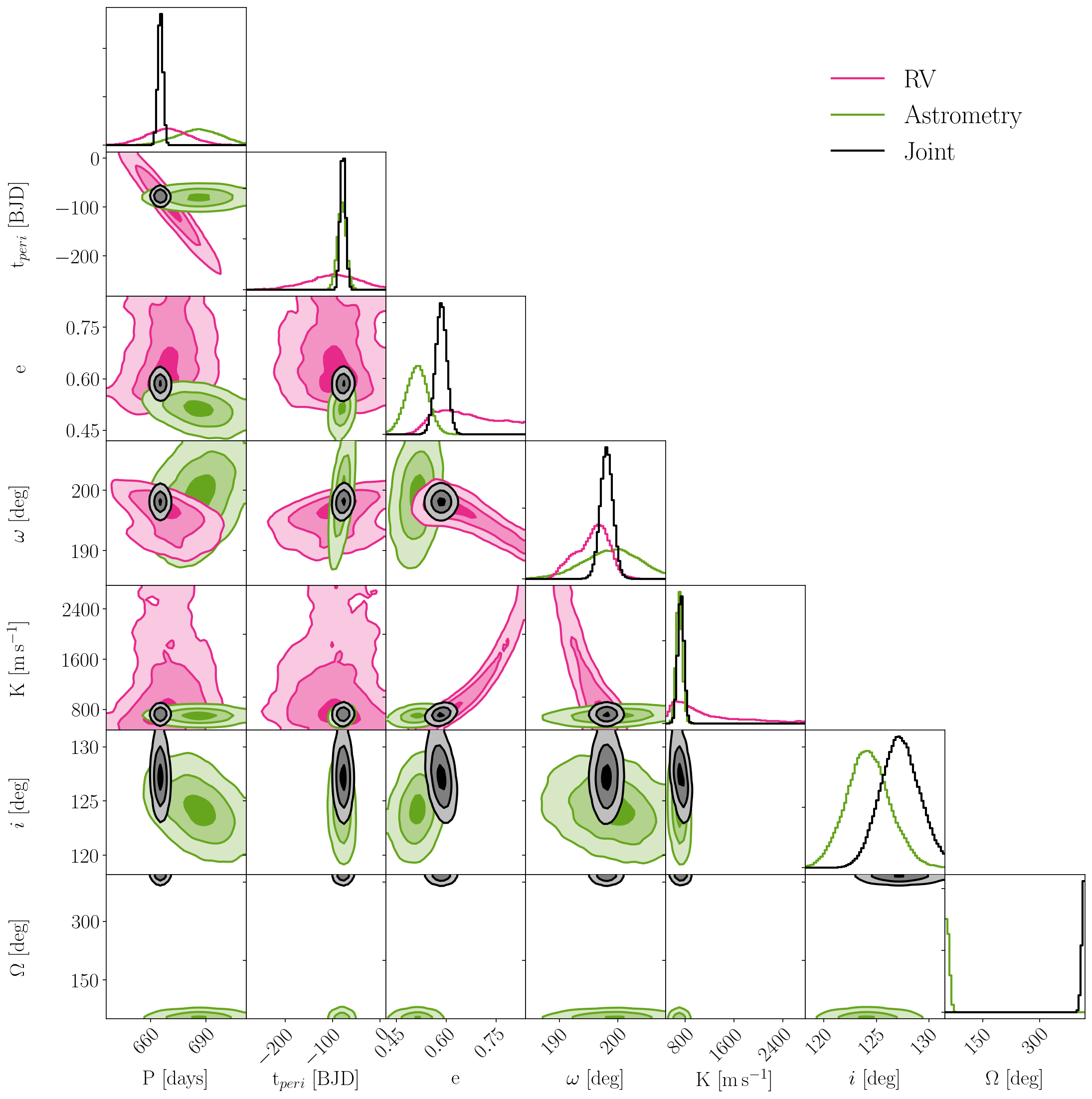}
    \caption{Same as Fig.~\ref{fig:corner7984}, but for LP 341-28.}
    \label{fig:corner4032}
\end{figure*}

\begin{figure*}
    \centering
    \includegraphics[width=18cm]{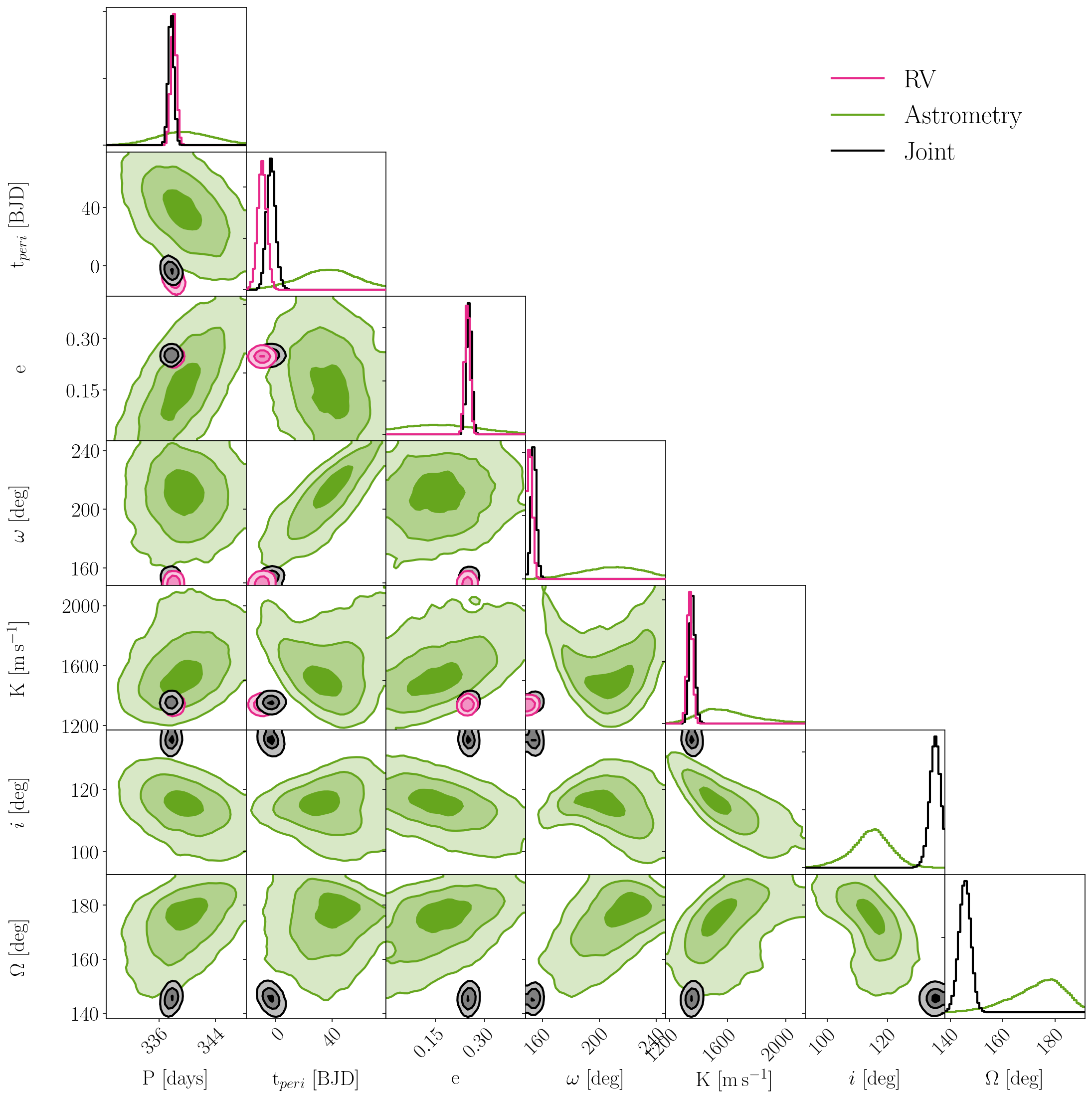}
    \caption{Same as Fig.~\ref{fig:corner7984}, but for LP 769-9.}
    \label{fig:corner11200}
\end{figure*}

\begin{figure*}
    \centering
    \includegraphics[width=18cm]{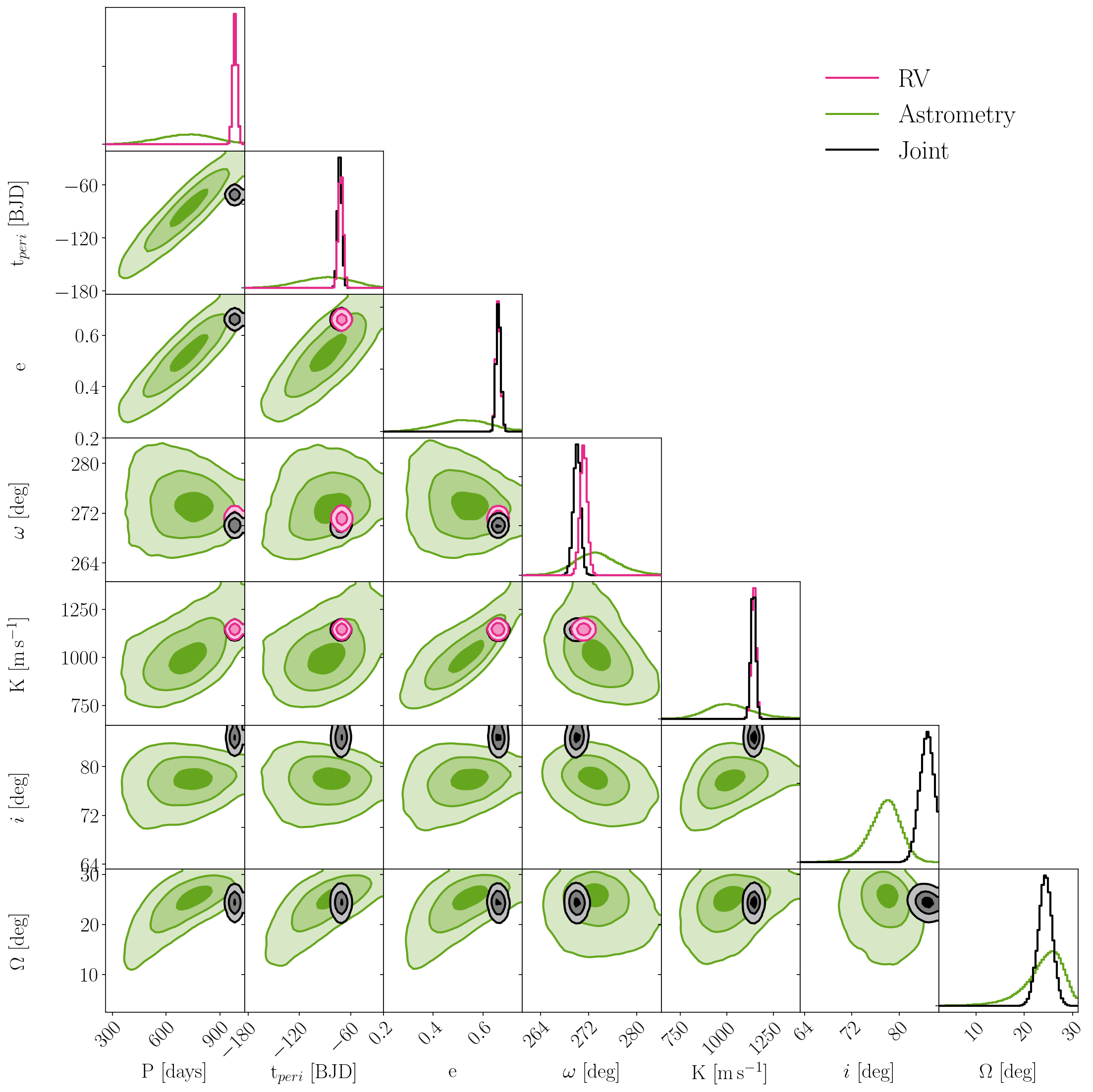}
    \caption{Same as Fig.~\ref{fig:corner7984}, but for HD 30246.}
    \label{fig:corner2048}
\end{figure*}

\begin{figure*}
    \centering
    \includegraphics[width=18cm]{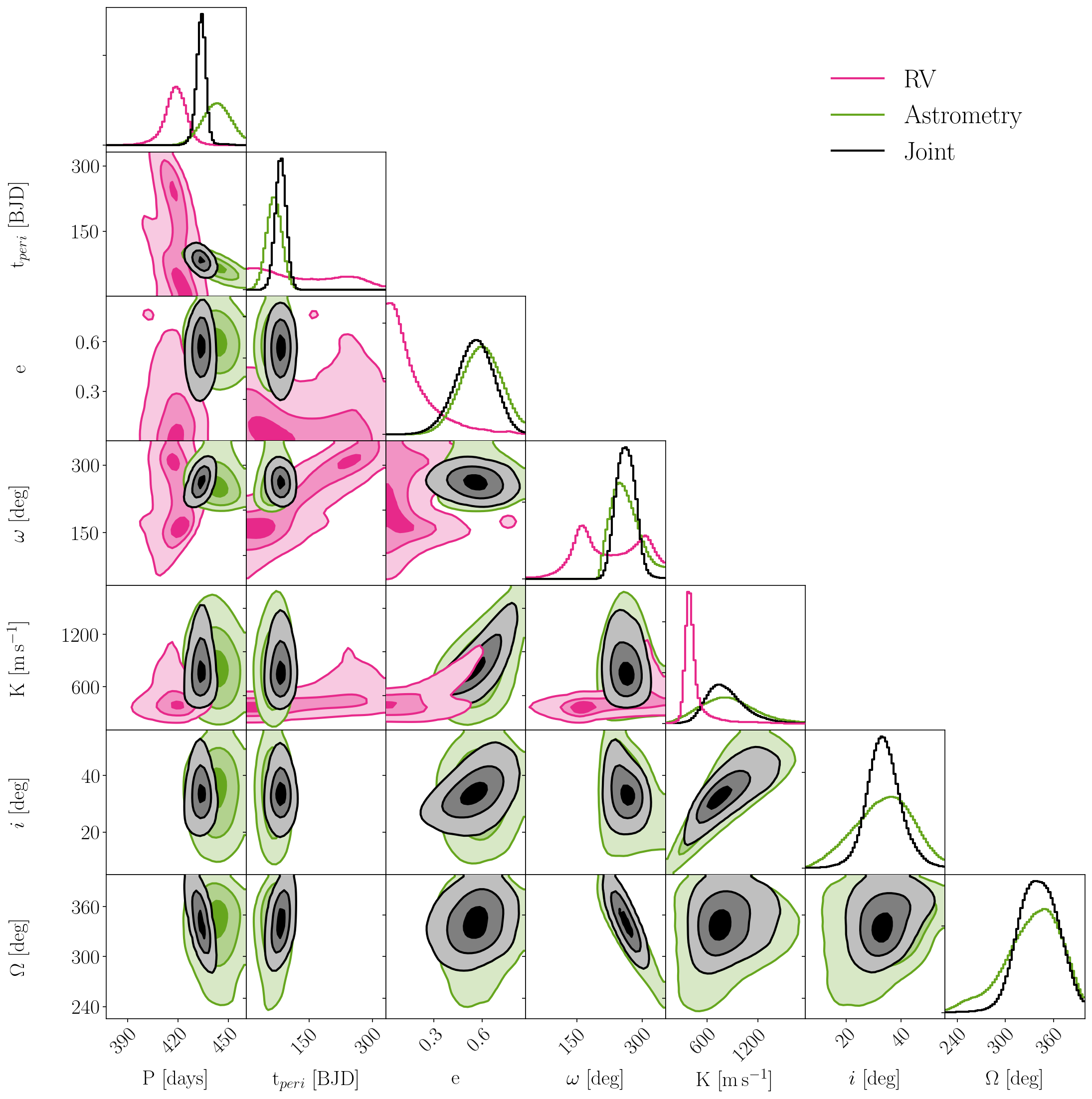}
    \caption{Same as Fig.~\ref{fig:corner7984}, but for 2M0809+07.}
    \label{fig:corner7408}
\end{figure*}

\end{appendix}
\end{document}